\documentclass[twocolumn,twocolappendix]{aastex63}
\usepackage{amsmath}

\newcommand\kms{km s$^{-1}$}

\newcommand\teff{$T_{\rm eff}$}
\newcommand\logg{$\log g$}

\newcommand\msun{M$_\odot$}

\begin{document}
\title{Double-lined spectroscopic binaries in the APOGEE DR16 and DR17 data}

\author[0000-0002-5365-1267]{Marina Kounkel}
\affil{Department of Physics and Astronomy, Vanderbilt University, VU Station 1807, Nashville, TN 37235, USA}
\affil{Department of Physics and Astronomy, Western Washington University, 516 High St, Bellingham, WA 98225}
\author[0000-0001-6914-7797]{Kevin R. Covey}
\affil{Department of Physics and Astronomy, Western Washington University, 516 High St, Bellingham, WA 98225}
\author[0000-0002-3481-9052]{Keivan G.\ Stassun}
\affil{Department of Physics and Astronomy, Vanderbilt University, VU Station 1807, Nashville, TN 37235, USA}
\author[0000-0003-0872-7098]{Adrian M. Price-Whelan}
\affil{Center for Computational Astrophysics, Flatiron Institute, 162 Fifth Avenue, New York, NY 10010, USA}
\author[0000-0002-9771-9622]{Jon Holtzman}
\affil{Astronomy Department, New Mexico State University, Las Cruces, NM 88003, USA}
\author[0000-0001-9984-0891]{Drew Chojnowski}
\affil{Department of Physics, Montana State University, P.O. Box 173840, Bozeman, MT 59717-3840}
\author[0000-0001-9330-5003]{Penélope Longa-Peña}
\affil{Centro de Astronomía (CITEVA), Universidad de Antofagasta, Av. Angamos 601, Antofagasta, Chile}
\author[0000-0001-8600-4798]{Carlos G. Rom\'an-Z\'u\~niga}
\affil{Universidad Nacional Autónoma de México, Instituto de Astronomía, AP 106, Ensenada 22800, BC, México}
\author[0000-0001-9797-5661]{Jesus Hernandez}
\affil{Universidad Nacional Autónoma de México, Instituto de Astronomía, AP 106, Ensenada 22800, BC, México}
\author[0000-0001-7351-6540]{Javier Serna}
\affil{Universidad Nacional Autónoma de México, Instituto de Astronomía, AP 106, Ensenada 22800, BC, México}
\author[0000-0003-3494-343X]{Carles Badenes}
\affil{PITT PACC, Department of Physics and Astronomy, University of Pittsburgh, Pittsburgh, PA 15260, USA}
\author[0000-0002-3657-0705]{Nathan De Lee}
\affil{Department of Physics, Geology, and Engineering Tech, Northern Kentucky University, Highland Heights, KY 41099, USA}
\author[0000-0003-2025-3147]{Steven Majewski}
\affil{Department of Astronomy, University of Virginia, P.O. Box 400325, Charlottesville, VA 22904, USA}
\author[0000-0003-1479-3059]{Guy S. Stringfellow}
\affil{Center for Astrophysics and Space Astronomy, University of Colorado, Boulder, CO 80309, USA}
\author[0000-0001-5253-1338]{Kaitlin M. Kratter}
\affil{Steward Observatory, University of Arizona, 933 N. Cherry Ave., Tucson, AZ 85721, USA}
\author[0000-0002-0870-6388]{Maxwell Moe}
\affil{Steward Observatory, University of Arizona, 933 N. Cherry Ave., Tucson, AZ 85721, USA}
\author[0000-0002-0740-8346]{Peter M. Frinchaboy}
\affil{Department of Physics \& Astronomy, Texas Christian University, Fort Worth, TX 76129, USA}
\author[0000-0002-1691-8217]{Rachael L. Beaton}
\affiliation{Department of Astrophysical Sciences, Princeton University, 4 Ivy Lane, Princeton, NJ~08544}
\affiliation{The Observatories of the Carnegie Institution for Science, 813 Santa Barbara St., Pasadena, CA~91101}
\author[0000-0003-3526-5052]{José G. Fernández-Trincado}
\affil{Instituto de Astronom\'ia y Ciencias Planetarias, Universidad de Atacama, Copayapu 485, Copiap\'o, Chile}
\author[0000-0001-9596-7983]{Suvrath Mahadevan}
\affil{Department of Astronomy and Astrophysics, The Pennsylvania State University, University Park, PA 16802}
\author[0000-0002-7064-099X]{Dante Minniti}
\affil{Departamento de Ciencias Fisicas, Universidad Andres Bello, Av. Fernandez Concha 700, Las Condes, Santiago, Chile}
\author[0000-0003-4573-6233]{Timothy C. Beers}
\affil{Department of Physics and JINA Center for the Evolution of the Elements, University of Notre Dame, Notre Dame, IN 46556, USA}
\author[0000-0001-7240-7449]{Donald P. Schneider}
\affil{Department of Astronomy and Astrophysics, The Pennsylvania State University, University Park, PA 16802}
\affil{Center for Exoplanets and Habitable Worlds, The Pennsylvania State University, University Park, PA 16802}
\author[0000-0003-1086-1579]{Rodolfo H. Barb\'a}
\affil{Departamento de Astronom\'{\i}a, Universidad de La Serena, Av. Juan Cisternas 1200 Norte, Chile}
\author[0000-0002-8725-1069]{Joel R. Brownstein}
\affil{Department of Physics and Astronomy, University of Utah, 115 S. 1400 E., Salt Lake City, UT 84112, USA}
\author[0000-0002-1693-2721]{Domingo Aníbal García-Hernández}
\affil{Instituto de Astrofísica de Canarias (IAC), E-38205 La Laguna, Tenerife, Spain}
\affil{Universidad de La Laguna (ULL), Departamento de Astrofísica, E-38206 La Laguna, Tenerife, Spain}
\author[0000-0002-2835-2556]{Kaike Pan}
\affil{Apache Point Observatory and New Mexico State
University, P.O. Box 59, Sunspot, NM, 88349-0059, USA}
\author[0000-0002-3601-133X]{Dmitry Bizyaev}
\affil{Apache Point Observatory and New Mexico State
University, P.O. Box 59, Sunspot, NM, 88349-0059, USA}
\affil{Sternberg Astronomical Institute, Moscow State
University, Moscow}

\email{marina.kounkel@vanderbilt.edu}
\begin{abstract}

APOGEE spectra offer $\lesssim$1 \kms\ precision in the measurement of stellar radial velocities (RVs). This holds even when multiple stars are captured in the same spectrum, as happens most commonly with double-lined spectroscopic binaries (SB2s), although random line of sight alignments of unrelated stars can also occur. We develop a code that autonomously identifies SB2s and higher order multiples in the APOGEE spectra, resulting in 7273 candidate SB2s, 813 SB3s, and 19 SB4s. We estimate the mass ratios of binaries, and for a subset of these systems with sufficient number of measurements we perform a complete orbital fit, confirming that most systems with period $<$10 days have circularized. Overall, we find a SB2 fraction ($F_{SB2}$) $\sim$3\% among main sequence dwarfs, and that there is not a significant trend in $F_{SB2}$ with temperature of a star. We are also able to recover a higher $F_{SB2}$ in sources with lower metallicity, however there are some observational biases. We also examine light curves from TESS to determine which of these spectroscopic binaries are also eclipsing. Such systems, particularly those that are also pre- and post-main sequence, are good candidates for a follow-up analysis to determine their masses and temperatures. 
\end{abstract}


\section{Introduction}

Binary systems are important in investigating a wide range of fundamental astrophysical problems \citep[e.g.,][]{duchene2013,moe2017}. Among them, short period binaries are of particular value, as it is possible to observe the two stars orbit each other multiple times to construct their full orbital solution to determine stellar properties. Specifically, spectroscopic binaries can be used to determine masses of individual stars.

Spectroscopic binaries (SBs) can be subdivided into different types. Single-line spectroscopic binaries (SB1s) have the flux in the spectrum dominated by the primary. They are identified through observing a periodic change in radial velocity (RV) of a particular source over time, and their initial detection  requires monitoring of a system for at least 2 epochs, preferably more. In contrast, double-lined spectroscopic binaries (SB2), or their higher order counterparts (SB3/SB4), have multiple stars of comparable brightness, such that signatures of both of them are captured in the same spectrum at the same time, producing several different characteristic velocities in a system. Such systems can be initially identified in just a single epoch, either through cross-correlating a spectrum against a single model template to identify distinct components, or through performing a multi-model fit of the spectrum.

While SBs are not uncommon, they are still relatively rare, with a typical SB multiplicity ratio of $\sim$5--10\% for studies that can achieve a sub-\kms\ precision in their RV measurements for G--M main sequence and pre-main sequence stars \citep[e.g.,][]{kounkel2019}. To build a sizable sample of candidates that can be used for rigorous statistics, it is necessary to obtain spectra of $>10^5-10^6$ sources, ideally over the course of several epochs. Studies of this nature are becoming increasingly common with large spectroscopic surveys, such as RAVE \citep[e.g.,][]{matijevic2010,birko2019}, LAMOST \citep[e.g.,][]{tian2020}, Gaia-ESO \citep[e.g.,][]{merle2017,merle2020}, GALAH \citep[e.g.,][]{traven2020}, and SDSS APOGEE \citep[e.g.,][]{fernandez2017,skinner2018,price-whelan2018,price-whelan2020}.

Nonetheless, analyzing such a large volume of data in a rigorous manner presents its own set of challenges. When searching for SBs in survey datasets, most studies focus on SB1s, as their identification is possible solely from the reduced catalog of RV measurements. Identification of higher order multiples requires custom data reduction and analysis pipelines. 

Previously, we developed a pipeline that autonomously identified SB2s in the APOGEE spectra \citep{kounkel2019} by deconvolving custom cross-correlation functions (CCFs). The \citet{kounkel2019} analysis, however, was restricted to APOGEE fields containing pre-main sequence stars and young clusters. In this work, we generalize the pipeline to be more effective on all stars that have been observed by the APOGEE survey. In Section \ref{sec:data} we describe the data used in the analysis. In Section \ref{sec:code} we discuss this automated code and the subsequent vetting of the identified SBs. In Section \ref{sec:results} we further derive various properties of the identified systems, such as their mass ratio. We also derive complete orbital solutions for the systems where there are sufficient data and also examine the light curves to flag the systems that may be eclipsing. In Section \ref{sec:discussion} we compare the derived sample to other catalogs of SBs and analyze the dependence of the derived SB2 fraction on various stellar properties. Finally, in Section \ref{sec:concl} we offer conclusions.

\section{Data}\label{sec:data}

Apache Point Observatory Galactic Evolution Experiment (APOGEE) is conducted with two high resolution spectrographs: one commissioned first on the 2.5-m Sloan Foundation telescope at the Apache Point Observatory (APO), and the second spectrograph (APOGEE-S) subsequently installed on the Ir\'en\'ee~du~Pont 2.5-meter telescope at Las Campanas Observatory (LCO) \citep{bowen1973,gunn2006,blanton2017}. Both spectrographs can observe up to 300 objects simultaneously, across 3$^\circ$ and 2$^\circ$ (in diameter) fields of view at APO and LCO, respectively. The spectrographs cover the spectral range of 1.51–1.7 $\mu$m with an average resolution of $R\sim$22,500 \citep{wilson2010,majewski2017,wilson2019}.

Throughout the survey, APOGEE has preferentially targeted red giants, however other stellar objects have been observed, often as part of dedicated goal or ancillary science programs \citep{zasowski2013,zasowski2017}. The data reduction pipeline is described in \citet{nidever2015,jonsson2020}, Holtzmann, J. et al, in prep. Multiple pipelines exist for deriving stellar parameters from the spectra, including ASPCAP \citep{garcia-perez2016}, which is the primary version of the parameters used by the survey, and it includes full chemical abundance determination. This pipeline identifies the best matching theoretical template spectrum through $\chi^2$ minimization in a multi-dimensional parameter space. Originally designed for measuring parameters for the giants, it has since been expanded to report parameters of all stars with \teff$<$8000 K. However, there are various systematic features in the reported parameter space due to theoretical templates not offering a perfect match to the real data. Another stellar parameter pipeline is APOGEE Net \citep{olney2020}, which is a neural network trained on empirical labels of stars, parameters of which have either been reliably measured by other spectral pipelines (i.e, giants), or those that could be estimated through phtometric relations (low mass dwarfs and pre-main sequence stars). This achieves a unified and self-consistent distribution of spectral parameters, with fewer obvious nonphysical systematic features. As such, it may offer a somewhat better \logg\ determination for the dwarfs.

The latest public data release is DR16 \citep{ahumada2020,jonsson2020}, covering observations through June of 2018. In this work, we analyze spectra from a proprietary data product including APOGEE observations taken through March of 2020, when both LCO and APO paused observations due to COVID-19. These observations were reduced with the DR16 pipeline and provided to the SDSS collaboration as an internal data release that contains spectra for 617,703 unique stars. SDSS-IV APOGEE data obtained after APO and LCO re-opened have only been processed by the DR17 iteration of the pipeline. The DR17 pipeline updated the procedures used to construct CCFs for each APOGEE spectrum, introducing slight differences between the DR16 and DR17 reductions that could not be reconciled in an autonomous manner. Thus, we treat these two data reductions separately, and present the treatment of the the DR17 data in Appendix \ref{sec:appendix}.

\section{Identification and characterization of SB2 candidates}\label{sec:code}

\subsection{Automated code}

In \citet{kounkel2019}, we developed a set of criteria that could be used to identify a stellar spectrum as an SB2 candidate based on the spectrum's CCF. There are some differences in the CCFs between this previous study and the work presented here. In \citet{kounkel2019}, we derived stellar parameters from scratch, as ASPCAP, particularly in the DR14 data release, did not have optimal parameters for the pre-main sequence stars (most notably, inaccurate \logg\ values). We then used a synthetic PHOENIX spectrum \citep{husser2013} that most closely matched the inferred parameters to compute a custom CCF extending to 200 \kms\ from the central peak, with 1 \kms\ linear spacing.

Extending our analysis from a few thousand stars in a few dozen fields to the entire APOGEE catalog made the calculation of new, independent CCFs computationally prohibitive. Therefore, we rely on the pre-computed CCFs that are made available on a visit level in all APOGEE apStar files \citep{nidever2015}. These CCFs are derived through masking out bad pixels, normalizing the spectra, and then performing a cross-correlation against a best matched synthetic template. In DR16 and earlier, this template was chosen from a small grid optimized for RV determination. In DR17, a synthetic spectrum is generated using the Doppler code \citep{doppler} through empirical data-driven models.

To make the resulting CCFs compatible with our existing analysis routines, we interpolated the APOGEE CCFs from their original exponential spacing to a linear grid with 1 \kms\ spacing. However, CCFs were extended from 200 to 400 \kms\ from the central peak, to ensure that multiple components in higher mass SB2s with very short periods and large RV separation could be recovered.

\begin{figure*}
\epsscale{1.1}
\plottwo{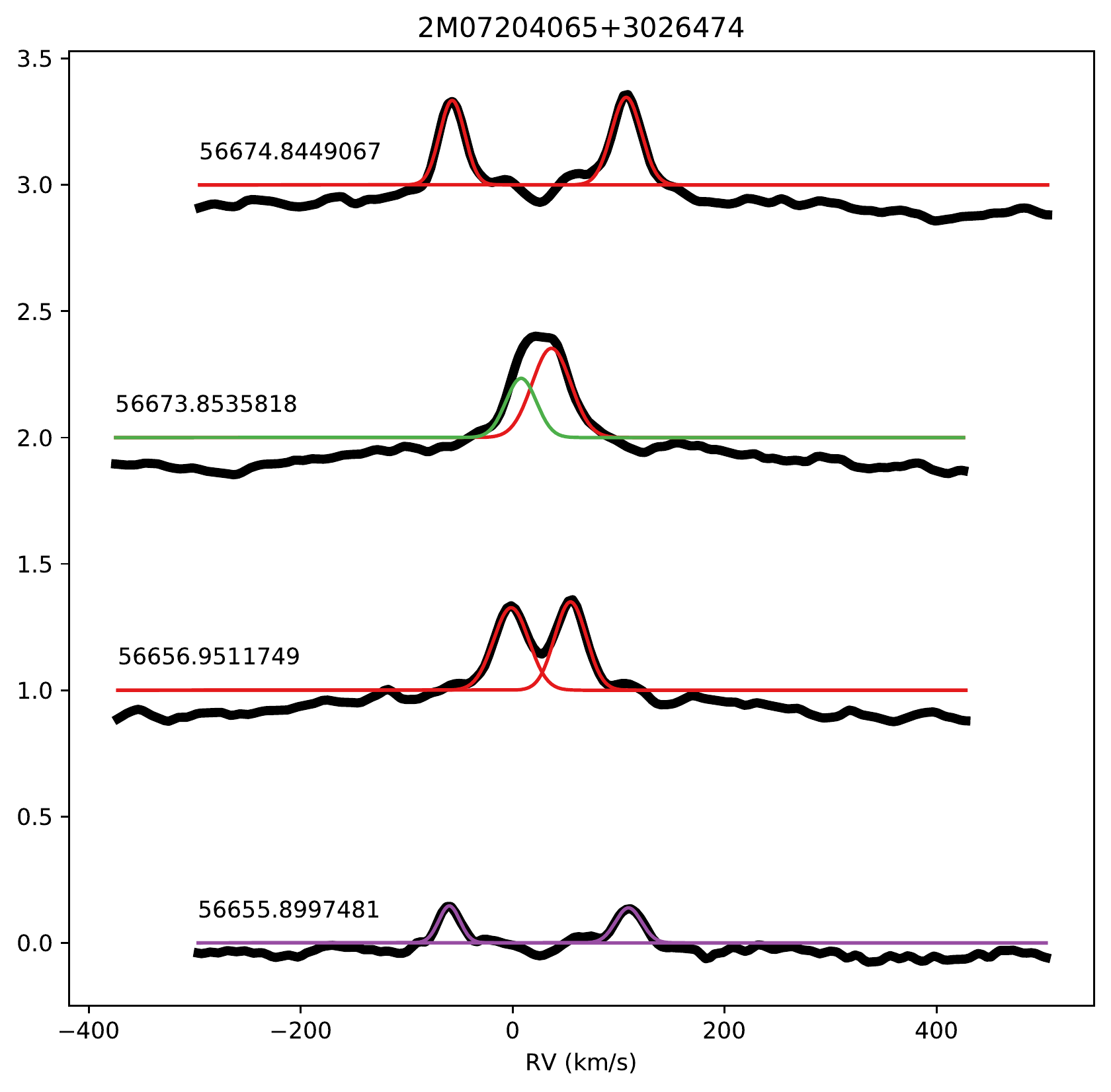}{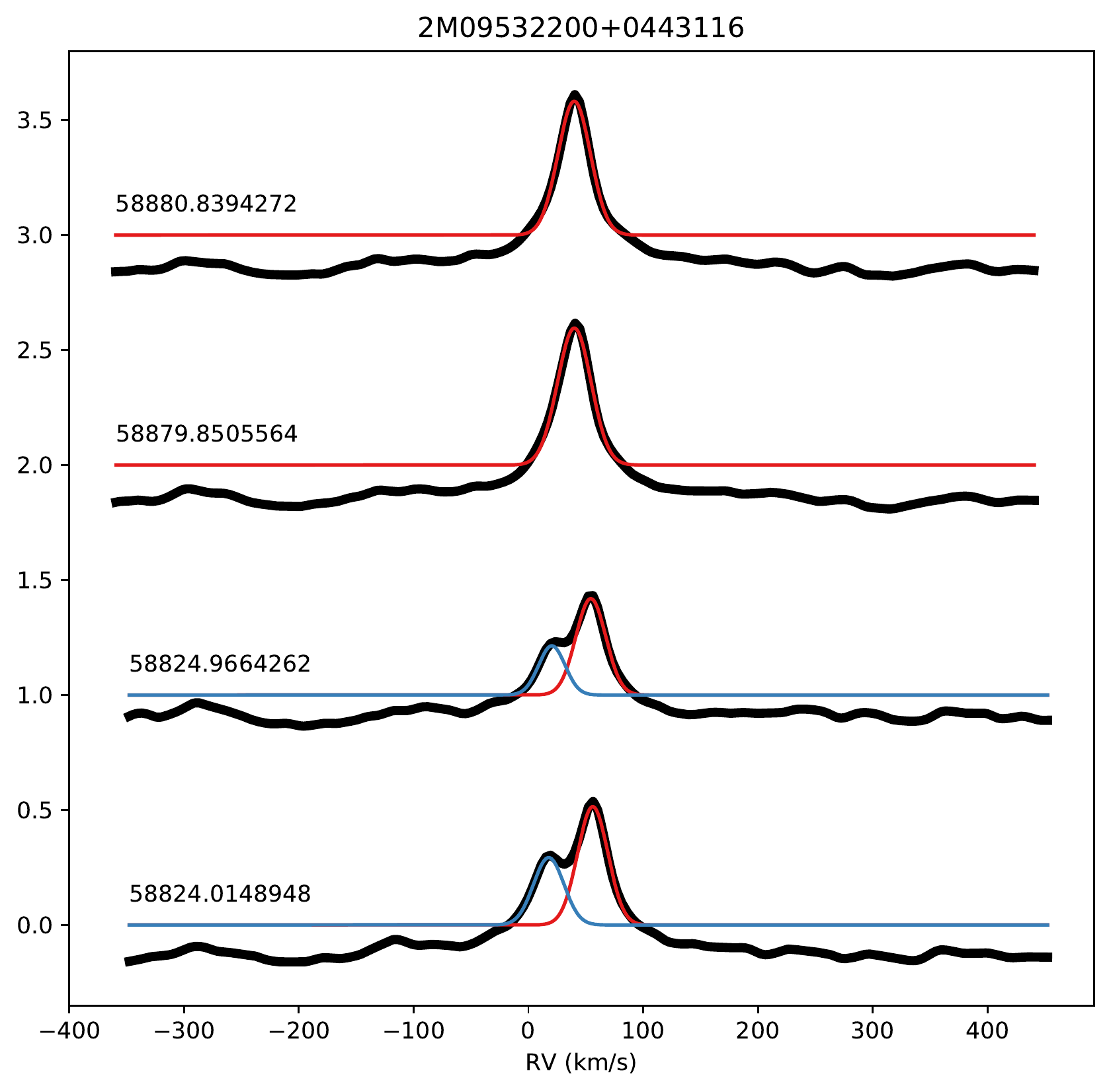}
\plottwo{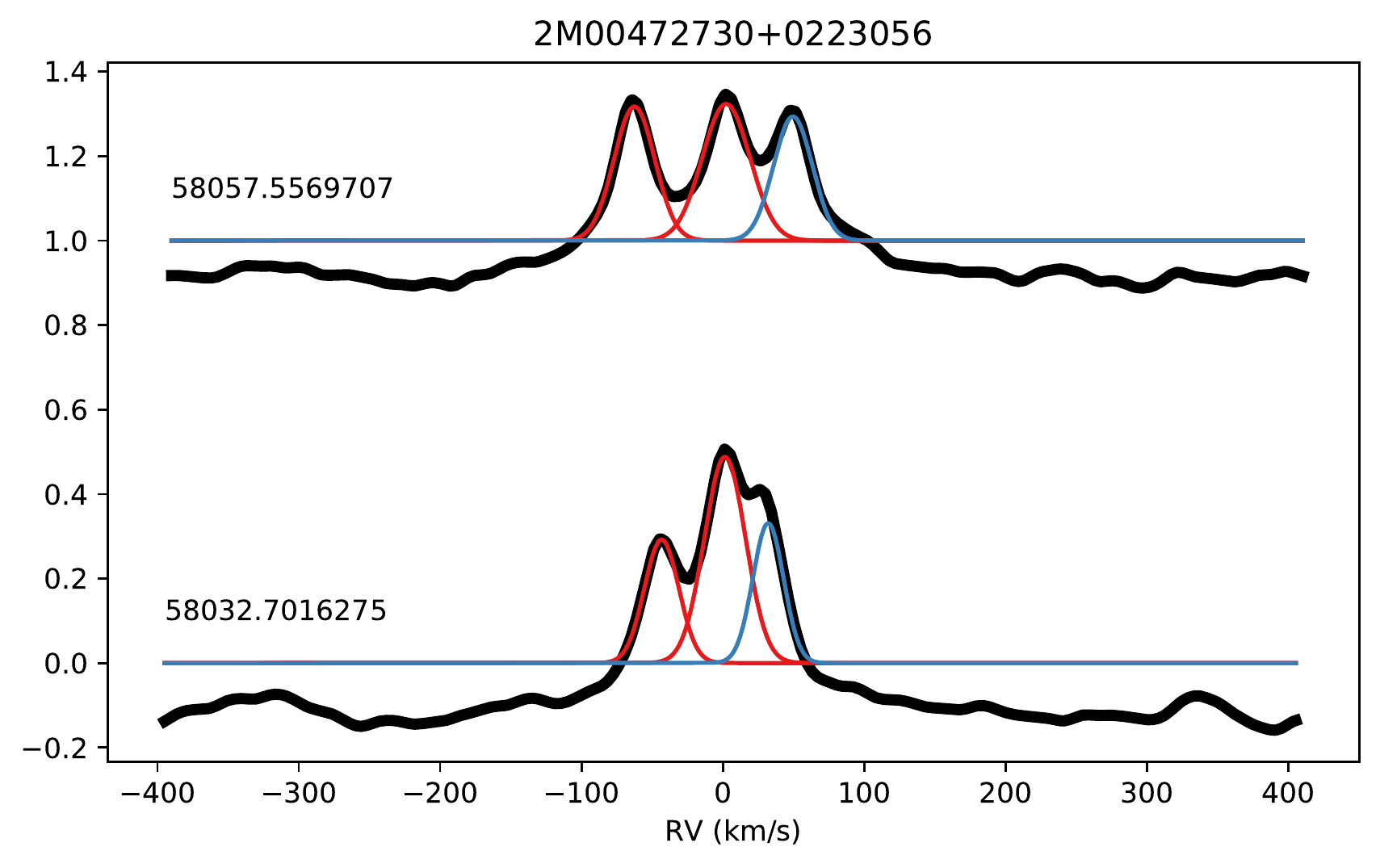}{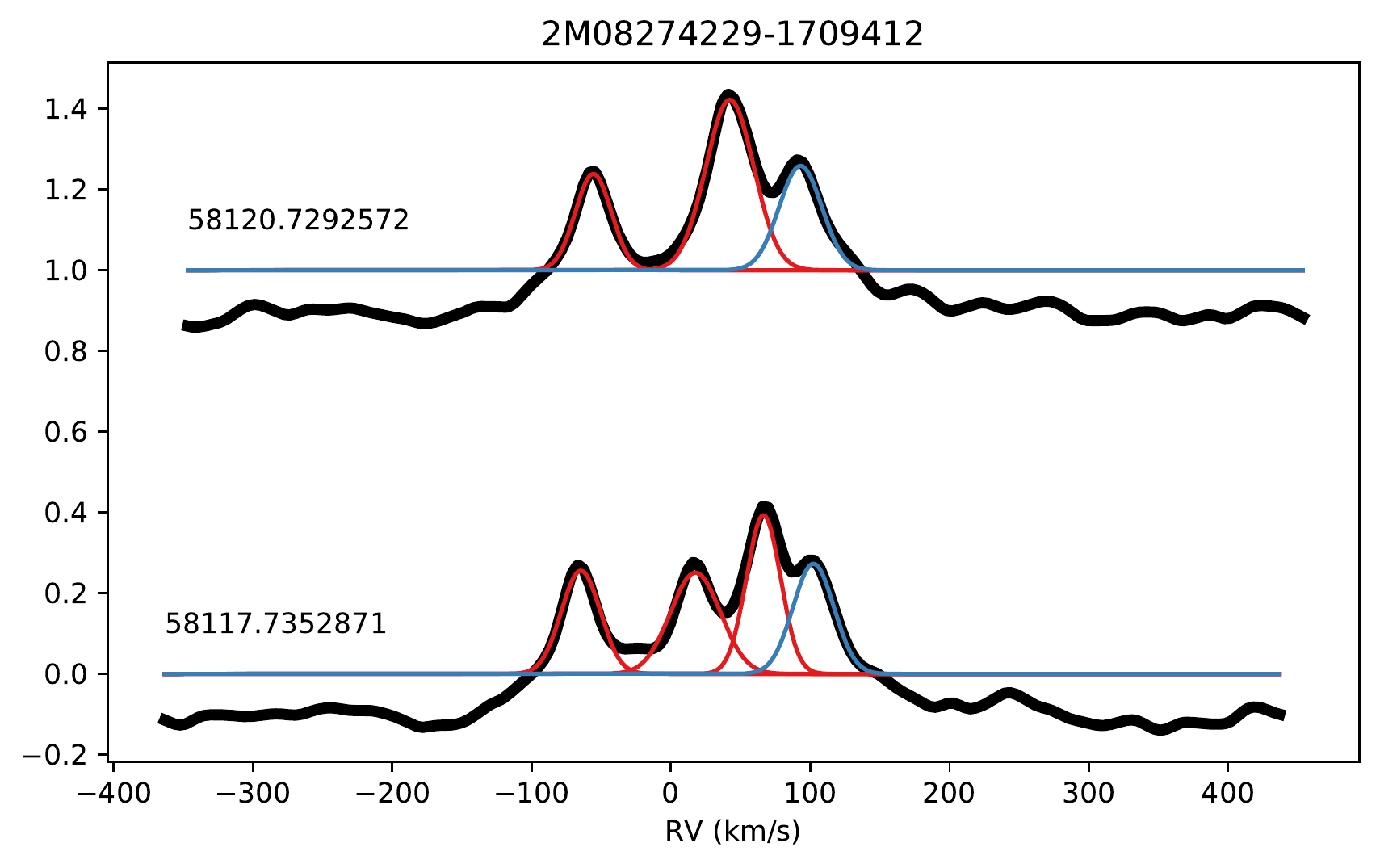}
\caption{Example of autonomously deconvolved CCFs. Black lines are the APOGEE CCFs. Red lines show components with Flag 4: primaries (in all cases), or widely separated secondaries. Blue lines show components with Flag 3: secondaries that fall within FWHM of the primary. Green lines show components with Flag 2 that failed the asymmetry check. Purple lines show components with Flag 1 that failed the amplitude test.
\label{fig:ccf}}
\end{figure*}

To identify peaks in each CCF that correspond with individual stars (Figure \ref{fig:ccf}), we first analyzed them with the autonomous Gaussian deconvolution python routine, GaussPy \citep{lindner2015}. To minimize GaussPy identifying noise in the CCF as a real signal, as well to prevent dominance of the non-Gaussian wings of the components, a continuum set to the median value of the CCF or 20\% of the peak (whichever was largest) was subtracted. Furthermore, the CCF was extended to zero at either end. In deconvolving the CCF, the $\log \alpha$ parameter was set to 1.5. The $\log \alpha$ parameter, defined in Equation 5 of \citet{lindner2015}, is the only variable used by GaussPy, specifying the characteristic scale of the recovered Gaussians by controlling the balance between real variance and noise.

After deconvolution, the identified Gaussian profiles were subjected to various tests, and assigned classification flags. Classification flag 1 was given to the components that were likely to be noisy or spurious, as identified from CCF peaks that had amplitudes of $<$0.15 or $>$3, or FWHM $<1$ or $>100$ \kms. Classification flag 2 was given to components that were likely falsely deconvolved into multiple pieces, identified as cases where secondary peaks lay within within 30 \kms\ of a primary peak that was too symmetric to plausibly contain multiple components within its width. Classification flag 3 was given to companions found inside a clearly asymmetric peak with a small RV separation, where both peaks are blended together. Finally, classification flag 4 was given to well resolved, well-separated components. See \citet{kounkel2019} for more details on the process of deriving and testing these various criteria.

Particularly for pre-main sequence and young stars, which were the focus of \citet{kounkel2019}, activity on the photosphere, such as that caused by spots \citep[e.g.,][]{hartmann1987a} can add structure to the resultant CCF, at the level indicated by flag 3. Therefore, \citet{kounkel2019} only deemed sources with multiple components with flag 4 as reliable SB2s. In contrast, more evolved main sequence and giant stars have less magnetic activity and less prominent spots, such that their CCFs are less likely to be affected by spot-induced structure. Therefore, here we consider components with both flags 3 \& 4 as likely SB2s. 

The deconvolution of multiple components is done on the per-epoch level. As binary stars are highly dynamical, the identified components may be flagged differently. For example, while two stars are well resolved, they may both be recovered with a classification flag of 4 -- on the other hand, if caught on the portion of their orbit where they would have similar RVs, they may still be recovered with a flag of 3 or 2, or, alternatively, they may blend into a single Gaussian profile (Figure \ref{fig:ccf}). If a particular epoch may have poor SNR, the amplitude of the Gaussian may decrease, and a system that would otherwise be recovered with flag 4 would instead have flag 1. As such, we require that a star is multiply deconvolved with either a flag 3 or 4 in just one single epoch to consider a star as an SB2 candidate. When several epochs are available, it is often likely for most of them to be deconvolved. However, it is not a requirement, as highly eccentric systems may have only a short window of time across their orbit during which a system may have sufficiently different RVs for both stars for the system to be resolved.

The python code that deconvolves the APOGEE CCFs and outputs a catalog of likely SB2s is made available on GitHub \citep{apogeesb2}\footnote{\url{https://github.com/mkounkel/apogeesb2}}. In the future, it will be integrated into \textit{Astra} - a pipeline under development for use by the SDSS-V survey, which continues to obtain stellar spectra with the APOGEE spectrographs. As such, beyond the value-added catalog presented in this work that is made available for the data observed through DR17, a list of sources that are identified as SB2s are likely going to be made available in subsequent data releases as well.

\begin{splitdeluxetable*}{ccccccccBccccccccBcccccccc}
\tabletypesize{\scriptsize}
\tablewidth{0pt}
\tablecaption{Properties of the identified SB2s and higher order multiples\label{tab:system}}
\tablehead{
\colhead{APOGEE} & \colhead{$\alpha$} & \colhead{$\delta$} & \colhead{SB$_N$} & \colhead{N} & \colhead{$q_w$\tablenotemark{$^a$}} & \colhead{$\gamma_w$\tablenotemark{$^a$}} & \colhead{$P$} & \colhead{$T_0$} & \colhead{$e$} & \colhead{$\omega$} & \colhead{$\gamma$\tablenotemark{$^b$}} & \colhead{$K_1$} & \colhead{$K_2$} & \colhead{$M_1\sin^3 i$} & \colhead{$M_2\sin^3 i$} & \colhead{$a\sin i$} & \colhead{Max $v$} & \colhead{Max $v_1$} & \colhead{Max $v_2$} & \colhead{Max $t$} & \colhead{LOS} & \colhead{TESS} & \colhead{TESS} \\
\colhead{ID} & \colhead{(deg.)} & \colhead{(deg.)} & \colhead{} & \colhead{epoch} & \colhead{} & \colhead{\kms} & \colhead{(day)} & \colhead{(JD)} & \colhead{} & \colhead{(deg.)} & \colhead{(\kms)} & \colhead{(\kms)} & \colhead{(\kms)} & \colhead{(\msun)} & \colhead{(\msun)} & \colhead{(au)} & \colhead{(\kms)}& \colhead{(\kms)}& \colhead{(\kms)}& \colhead{(day)}& \colhead{(?)} & \colhead{variable?\tablenotemark{$^c$}} & \colhead{period (day)}
}
\startdata
 2M09304810+2712390 & 142.700448 & 27.210835 & 2 & 29 & 0.95$\pm$0.01 & 8.98$\pm$0.47 & 4.205994 $\pm$1.2741106E-5 & 2457677.8 $\pm$ 0.06 & 0.006$\pm$ 0.002 & 165$\pm$ 5 & 9.04$\pm$0.12 & 87.3$\pm$0.2 & 91.1$\pm$ 0.3 & 1.263 $\pm$0.008 & 1.210$\pm$0.007 & 0.06895$\pm$0.00014 & 180.3 & 177.8 & 183.5 & 1911.7 & & d & 4.212\\
\enddata
\tablenotetext{}{Only a portion shown here. Full table with is available in an electronic form.}
\tablenotetext{}{$\alpha$=RA, $\delta$=Dec., SB$_N$=number of deconvolved components for a source, $q$=mass ratio, $\gamma$=barycenter velocity,$P$=period, $T_0$=time of periastron passage, $e$=eccentricity, $\omega$=Longitude of periastron, $K_{1,2}$=semiamplitude of the velocity, $M_{1,2}\sin^3 i$=inclination dependent mass of the star, $a\sin i$=inclination dependent semimajor axis, Max $v$=maximum observed separation in RV between the primary and the secondary in any epoch, Max $v_{1,2}$=observed amplitude of variation in RV of the component across all of the available data, Max $t$=maximum temporal baseline, LOS= line of sight coincidence}
\tablenotetext{a}{From Wilson plot}
\tablenotetext{b}{From orbital fit; $P$=period, $T_0$=time of periastron passage, $e$=eccentricity, $\omega$=Longitude of periastron, $\gamma$=barycenter velocity, $K_1$=Semiamplitude of the velocity for primary, $K_2$=Semiamplitude of the velocity for secondary}
\tablenotetext{c}{f=in TESS footprint; v=variable; d=detached eclipsing}
\end{splitdeluxetable*}

\subsection{Validation}

\begin{figure*}
\epsscale{1.1}
\plotone{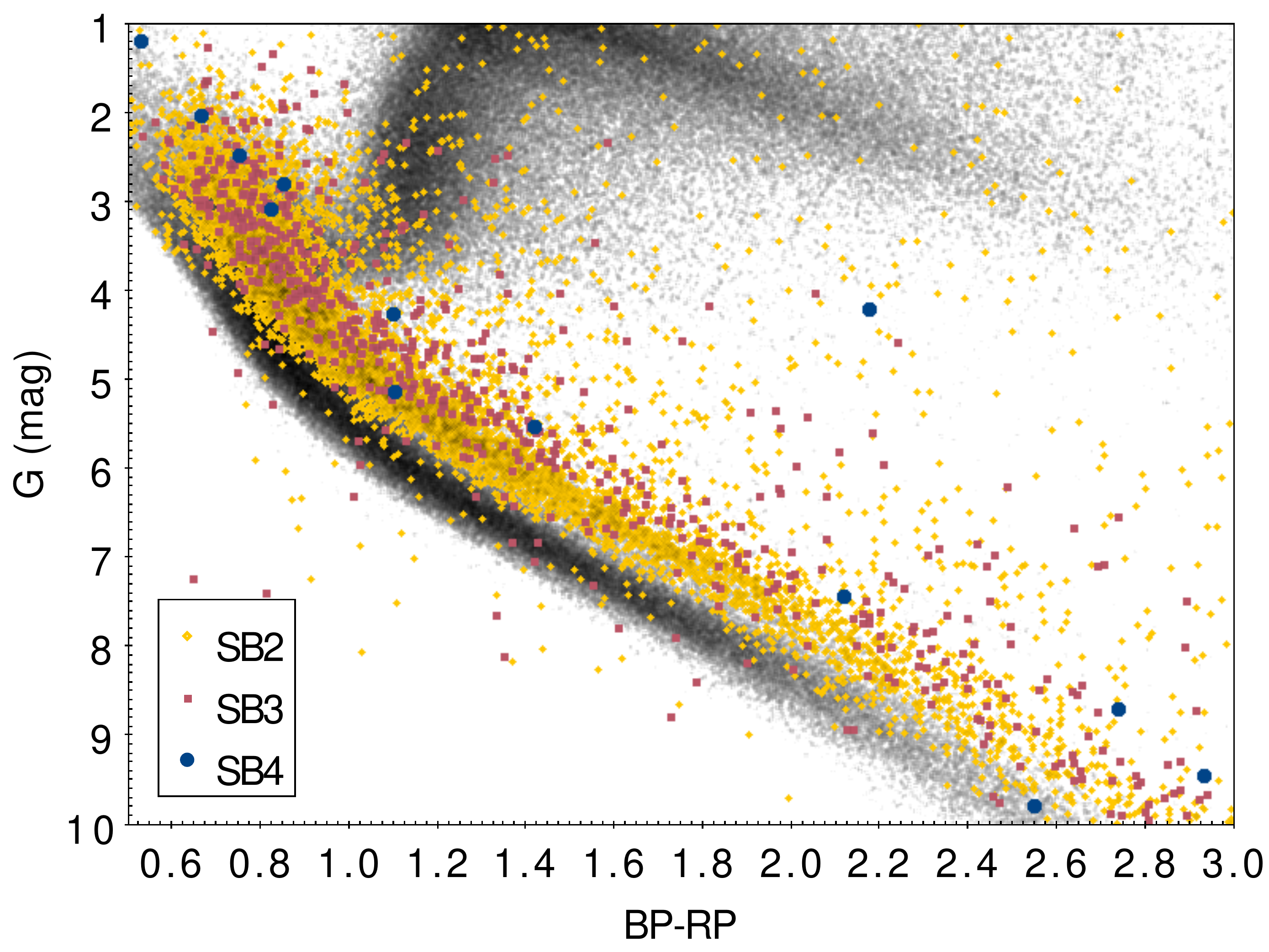}
\plottwo{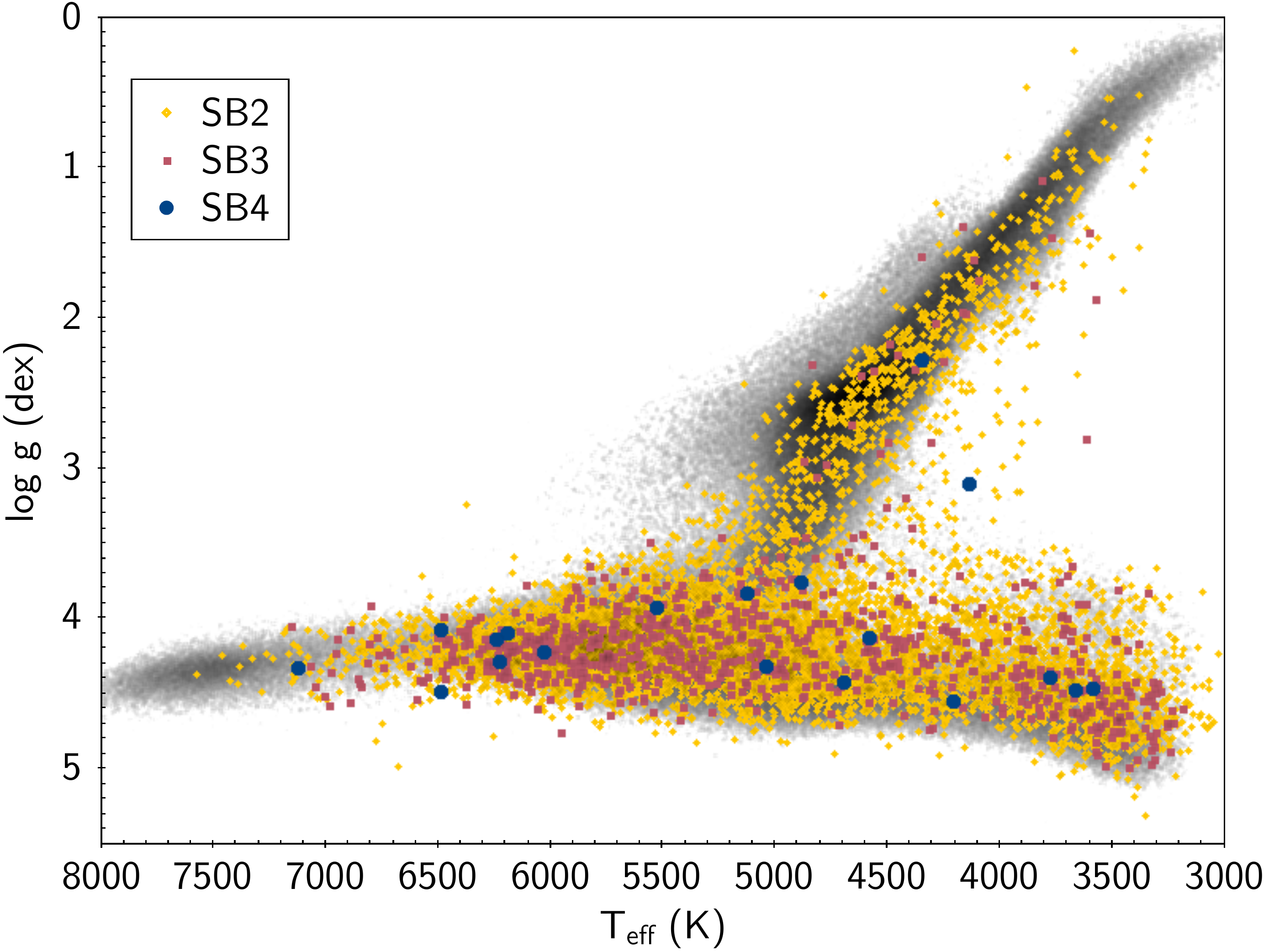}{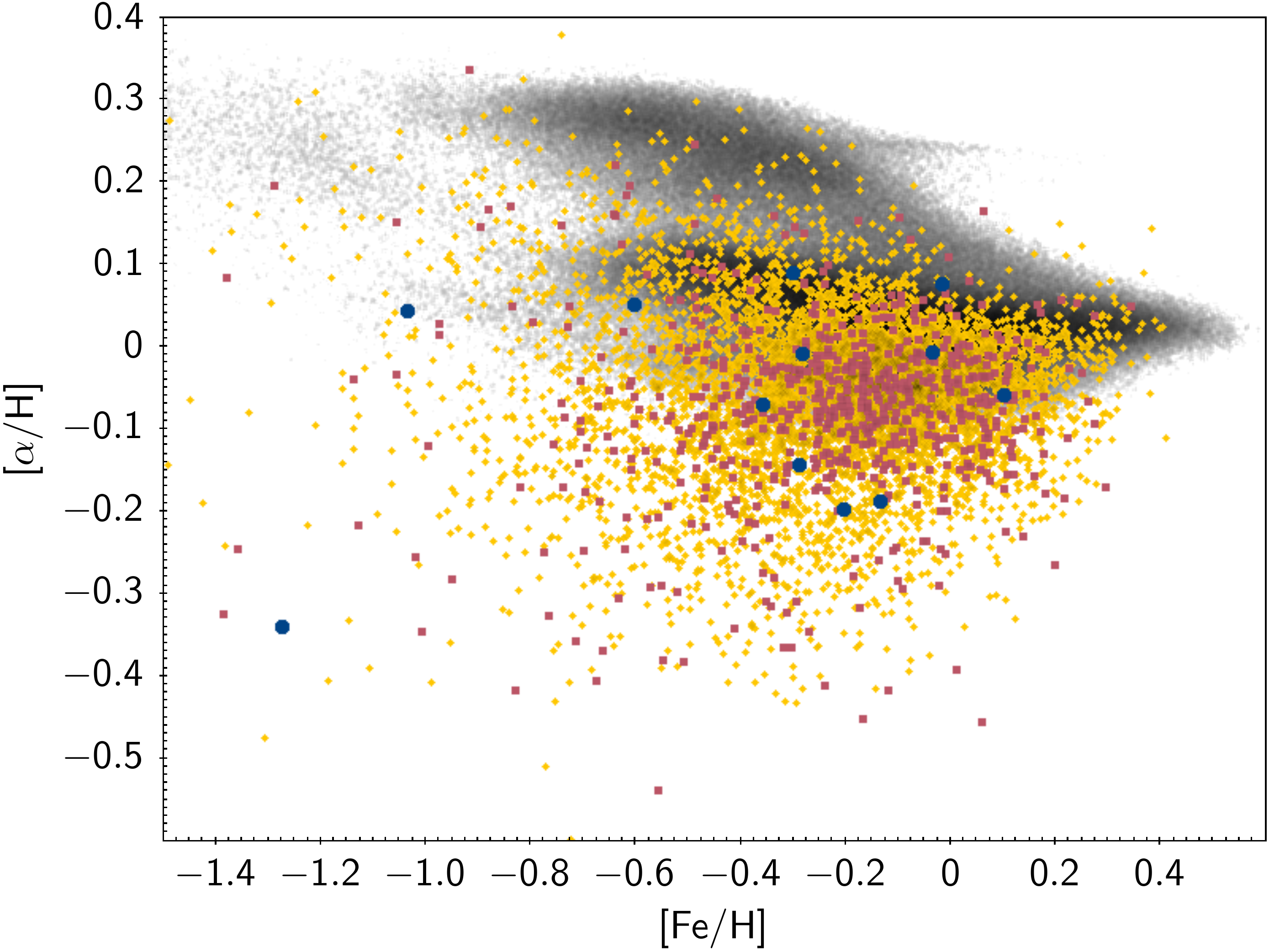}
\caption{Top: HR diagram of the spectroscopic binaries in comparison to the full sample. Note that SB2s tend to be located along the binary sequence, and SB3s and SB4s tend to be brighter than SB2s. The sources found below main sequence tend to have large uncertainties in the parallax, with $\sigma_\pi>0.2$ mas. Bottom left: \teff-\logg\ distribution using APOGEE Net parameters \citep{olney2020}. Bottom right: [Fe/H] vs [$\alpha$/H] distribution from ASPCAP DR17 \citep[Holtzmann J. et al in prep]{garcia-perez2016}.
\label{fig:sbn}}
\end{figure*}

\begin{deluxetable}{ll}
\tabletypesize{\scriptsize}
\tablewidth{0pt}
\tablecaption{Statistics in the sample\label{tab:stats}}
\tablehead{
\colhead{Property} &\colhead{Number}
}
\startdata
\multicolumn{2}{c}{Raw Deconvolution} \\
\hline
Number of candidate SBs & 8538 \\
Number of visits for all candidate SBs & 34903 \\
Number of visits with second component with Flag$\geq$3 & 20224 \\
Number of visits with third component with Flag$\geq$3 & 592 \\
Number of visits with fourth component with Flag$\geq$3 & 3 \\
\hline\hline
\multicolumn{2}{c}{After visual vetting} \\
\hline
Number of vetted SBs & 8105 \\
Number of SB2s & 7273 \\
Number of SB3s & 813 \\
Number of SB4s & 19 \\
\hline
Number of visits for all vetted SBs & 32642 \\
Number of visits with two visible components & 28277 \\
Number of visits with three visible components & 2169 \\
Number of visits with four visible components & 40 \\
\hline
Number of SBs with 1 epoch & 1842 \\
Number of SBs with 2 epochs & 2011 \\
Number of SBs with 3-5 epochs & 2679 \\
Number of SBs with $>$5 epochs & 1573 \\
\hline\hline
\multicolumn{2}{c}{Orbits} \\
\hline
Number of systems with $q$ and $\gamma$ from Wilson plot & 4658 \\
Number of systems with full orbits & 320 \\
\hline\hline
\multicolumn{2}{c}{Light curves} \\
\hline
Number of SBs in TESS footprint & 5485 \\
Number of SBs with periodic light curves & 1504 \\
Light curves of detached eclipsing binaries & 369 \\
EBs with matching orbital periods to stand-alone RV fit & 82\\
\hline
\enddata
\end{deluxetable}

While the automated code is useful for identifying SB2s in bulk and estimating preliminary RVs for their individual components, autonomous deconvolution will nonetheless sometimes report spurious peaks. Alternatively, deconvolution may fail to detect each component at all epochs, such as by fitting only a single Gaussian to two closely spaced components, or missing the second peak entirely. Futhermore, although components with flags 3 \& 4 tend to be robust, there is some ambiguity for components with flags 1 \& 2 -- while most of the sources with flags 1 and 2 are false positives, if a firmer detection exists in other epochs, some of them do trace real stellar counterparts, but require manual vetting to confirm.

Thus, we visually examined all 34,903 CCFs for the suspected SB2s identified by our routines in the mid-2020 APOGEE internal data release (Table \ref{tab:stats}). We used a custom Python tool to interactively refit the CCF by changing the continuum, and specifying the number of Gaussians to fit as well as their initial parameter estimates. Through this, we manually curated the catalog, paying particular attention to the evolution of the shape of the CCF between epochs to ensure self-consistency and an optimal fit for each source. Whenever necessary, missing components at a given epoch were added, and components that were suspected to be noise were removed.

In this process, some sources were judiciously discarded from the catalog of likely SB2s if the CCF was split into multiple components due to either the width or noise of the primary peak in the CCF, but was not algorithmically assigned classification flags 2 or 3. Similarly, sources with irregular CCFs that were difficult to fit through Gaussian deconvolution were also rejected if other epochs did not support that the source was a likely binary. In the latter case, this signature could be related to the photospheric spots originally targeted by classification flag 3, as such sources were prominent primarily in star forming regions and young clusters previously analyzed in \citet{kounkel2019}.

\section{Results}\label{sec:results}

\begin{deluxetable*}{cccccc}
\tabletypesize{\scriptsize}
\tablewidth{0pt}
\tablecaption{Vetted radial velocities of the individual components\label{tab:epoch}}
\tablehead{
\colhead{APOGEE} &\colhead{HJD} &\colhead{$v_1$} & \colhead{$v_2$} & \colhead{$v_3$}& \colhead{$v_4$}\\
\colhead{ID} &\colhead{(day)} &\colhead{(\kms)} & \colhead{(\kms)} & \colhead{(\kms)}& \colhead{(\kms)}
}
\startdata
2M19534348+2804599 & 56936.584 & 95.7$\pm$1.4 & -38.6$\pm$5.0 & 43.8$\pm$6.0 & -73.6$\pm$3.2\\
\enddata
\tablenotetext{}{Only a portion shown here. Full table with is available in an electronic form.}
\end{deluxetable*}

\begin{figure*}
\epsscale{1.1}
\plottwo{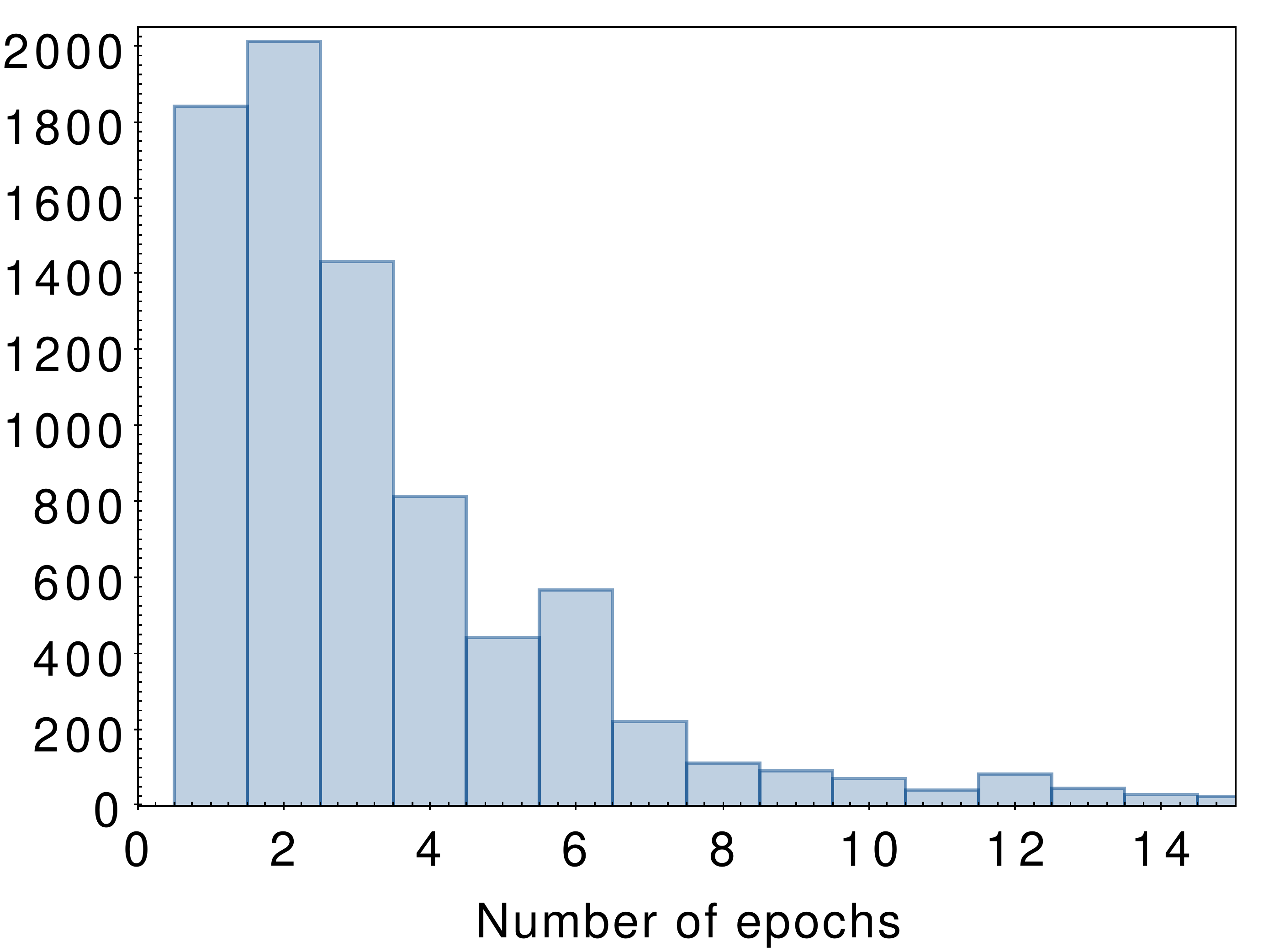}{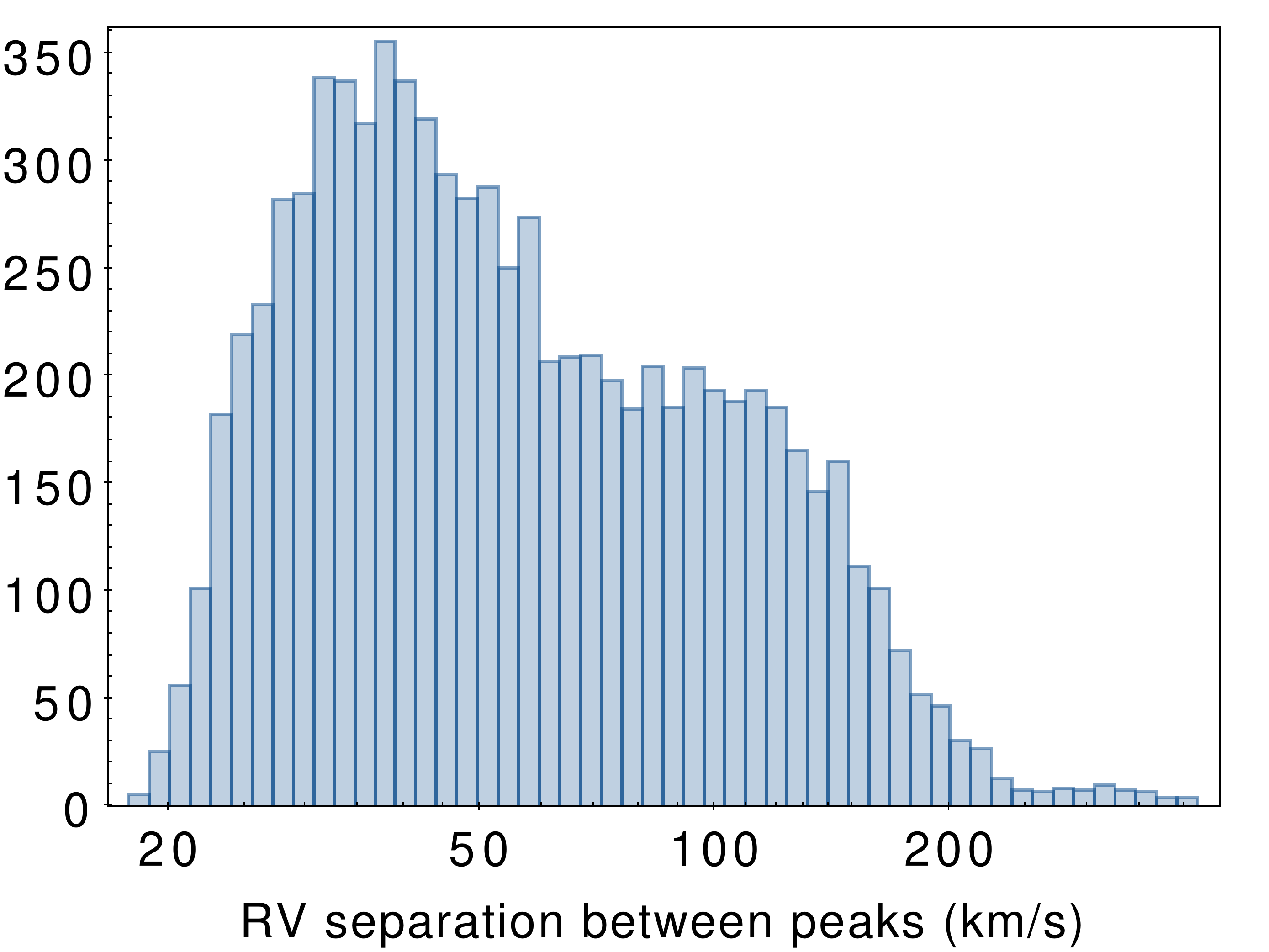}
\caption{Left: Distribution of the number of epochs available for the SB2s and higher order systems. Right: Distribution of the maximum separation between velocities of the primary and the secondary.
\label{fig:data}}
\end{figure*}

In total, 7273 sources were identified as SB2s, 813 were identified as SB3s, and 19 as SB4s (Figure \ref{fig:sbn}). Most sources have only been observed for 1--2 epochs, and the minimum separation between multiple peaks is $>$20 \kms\ (Figure \ref{fig:data}). Comparing the automatic and vetted catalogs, we estimate a false positive fraction of $\sim$3\% for components with flag 3 or 4, and $\sim$6\% for any secondary component to be false positive with any flag (in sources where a firm detection exists in other epochs). Similarly, based on the epochs where a second component was not identified with a flag 3 or 4, we estimate the completeness of the autonomous deconvolution of $\sim$80\%. Completeness increases to 90\% for a secondary component that can be identified through visual examination to be deconvolved regardless of a flag. If a source has multiple epochs, there is a greater probability that it would be flagged as a multiple in some of them and thus be a part of the catalog regardless. But, if a source was observed only once then it is possible that it would be missed. Note that it does not take into account incompleteness due to the construction of the CCF itself, as the template that is used to make a CCF matters strongly in whether a secondary is detectable in the first place. For example, in the sample from \citet{kounkel2019} using independently derived stellar parameters and custom CCFs, there are almost twice as many SB2s compared to what is presented here for the same sources.

Examining the HR diagram of the identified multiples in comparison to the full APOGEE catalog, they preferentially lie on the binary and tertiary sequences of dwarf stars (Figure \ref{fig:sbn}). A small but sizable population ($\sim$10\%) appear to be red giants -- while many are associated with true multiple systems, a considerable number of them are likely to be line of sight contaminants (Section \ref{sec:los}). Such contaminants are not counted towards false positive statistics as they do nonetheless have a spectrum composed of multiple stars.

\subsection{Line of sight systems} \label{sec:los}

The majority of identified systems are bona fide binary or tertiary stars. There are, however, a number of CCFs with two or more distinct components that are chance alignment of unrelated stars along the same line of sight captured in the same spectrum. As the fiber diameter is 2'' for the spectrograph at APO and 1.3'' for the spectrograph at LCO, sources with the apparent separation smaller than that that may be affected as long as they have comparable flux.

Such systems could be identified in a few different ways. 

\begin{itemize}
\item Most apparently, the velocity separation of the components is substantial ($>$40 \kms\ ), but doesn't significantly change over several epochs. For a system with a $\leq$ 10 \msun\ stellar primary and 1 \msun\ secondary, for example, orbits that produce such large velocity separations would have periods of $\sim$1 day, such that stationary velocities over timescales of several days to years prove that the components contributing to the spectrum are not interacting gravitationally with one another. 
\item Such line of sight systems can also be detected in a single epoch if the components have a very large RV separation. Difference in RV of greater than 200 \kms\ is not unusual, occasionally exceeding 400 \kms. This is unlikely even in contact binaries, which could have their RV separation range up to $\sim$200 \kms\ in dwarfs and up to $\sim$100 \kms\ in giants \citep{badenes2018}.
\end{itemize}

Following their identification, the line of sight systems tend to have particular properties in terms of their spatial locations and evolutionary status.

\begin{itemize}
\item Such systems are primarily concentrated towards the Galactic center, and to a lesser extent along the Galactic plane. Along these lines of sight the crowding is the highest, resulting in a larger number of chance alignments.
\item Such systems tend to consist of red giants by a factor of 2/3 in comparison to the dwarfs. While bona fide SB2s among red giants are not unprecedented, they are nonetheless rare, requiring a brightness ratio on the order of 1 between two components. As such, this is easier to achieve for unrelated stars, such as because they are found at different distances and because of the larger potential pool of stars. Because of their evolutionary status, they tend to have very narrow CCF profiles, which are visually distinct from the CCF profiles of dwarfs. 
\end{itemize}

\begin{figure*}
\epsscale{1.1}
		\gridline{\fig{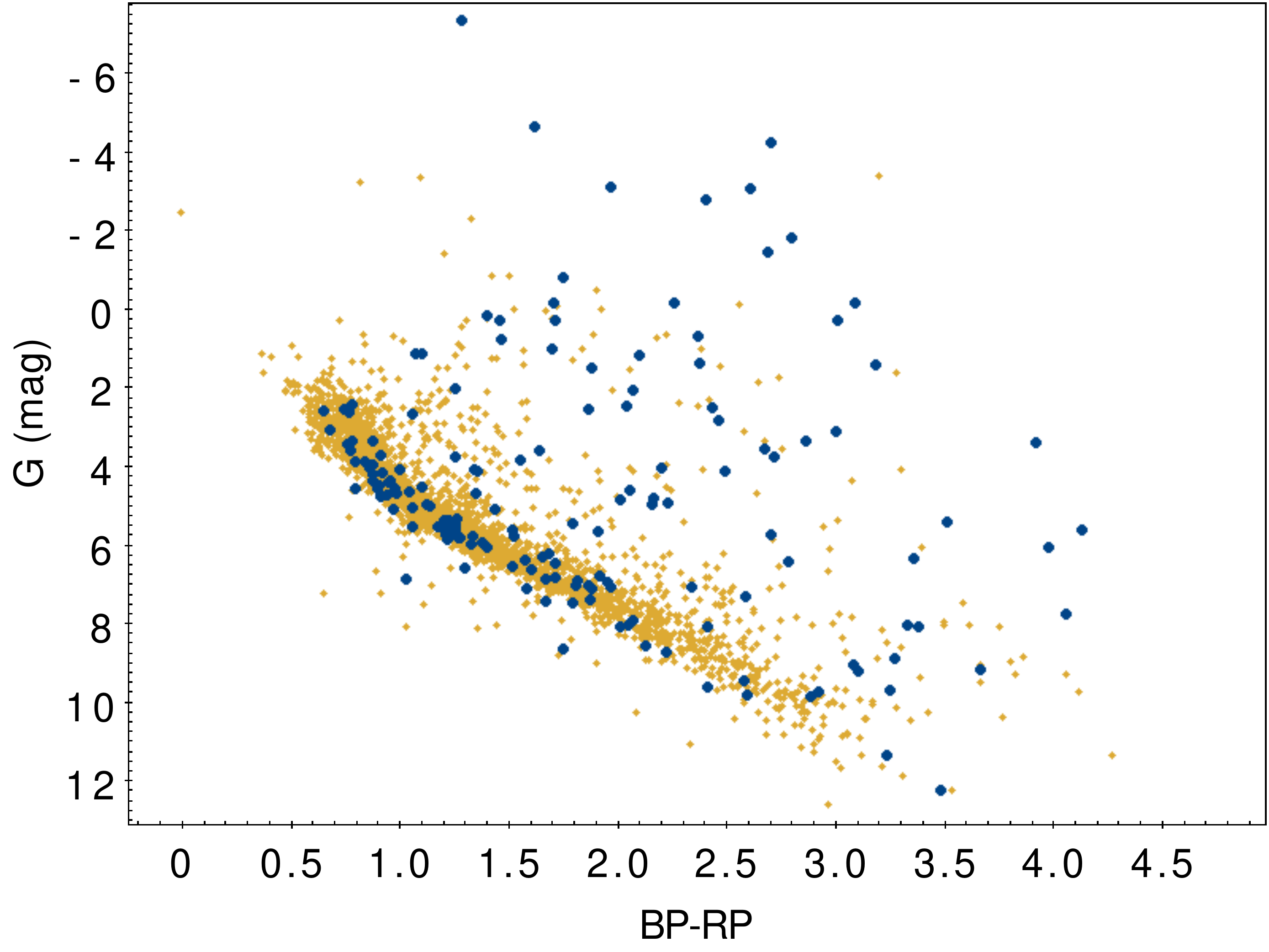}{0.4\textwidth}{}
 \fig{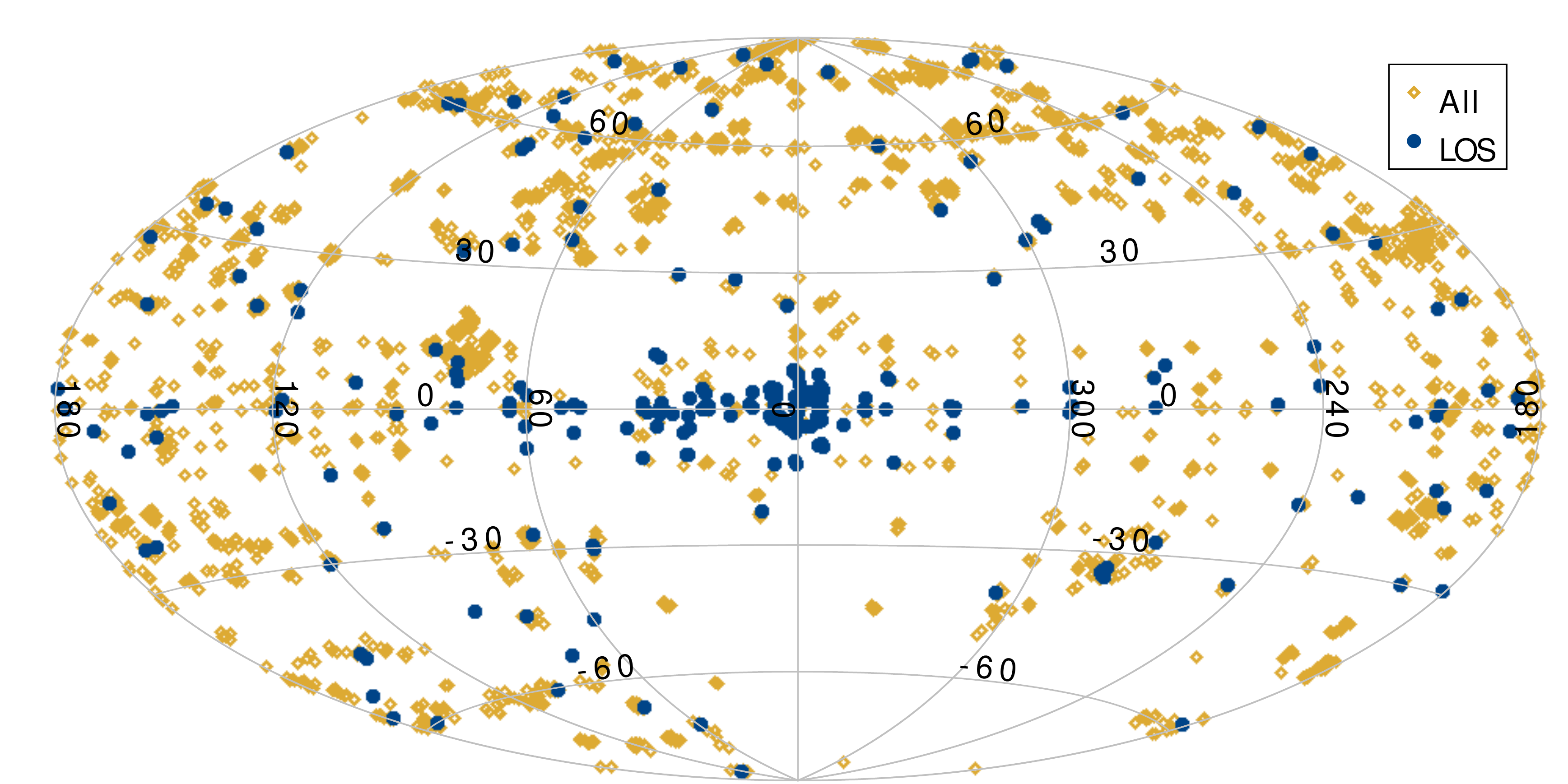}{0.6\textwidth}{}
 }
\caption{Distribution of the systems that are likely line of sight coincidences (LOS, blue) in comparison to the full sample (yellow, minimum of 3 epochs separated by at least 30 days). Left: distribution of sources on the HR diagram. Right: Position of the sources in the plane of the sky, in the galactic coordinates. Note that LOSs tend to be located towards the Galactic center, and they tend to be red giants. The contamination fraction of these LOSs can reach as much as 2--8\% of the full observed sample in these fields towards the Galactic center due to high stellar density.
\label{fig:los}}
\end{figure*}

Out of 2978 systems that have been observed for a minimum of 3 epochs, with a minimum temporal baseline of at least 30 days, there are 286 systems that are likely line of sight coincidences, defined as having the total RV change for each component of less than 5 \kms\ (Figure \ref{fig:los}). Towards the Galactic center, the fraction of such systems can reach as much as 6--8\% out of the total number of stars observed by APOGEE in those fields, up to 2\% along the disk within the inner 30$^\circ$, but, otherwise, line of sight systems are uncommon. Of these 286 systems, 149 of them can be cross-matched to multiple sources within Gaia EDR3 within 2'' search radius. The remaining systems may be too close on the sky to be resolved by Gaia in the current data release.

\subsection{Mass ratio}\label{sec:q}

\begin{figure*}
\epsscale{1.1}
		\gridline{\fig{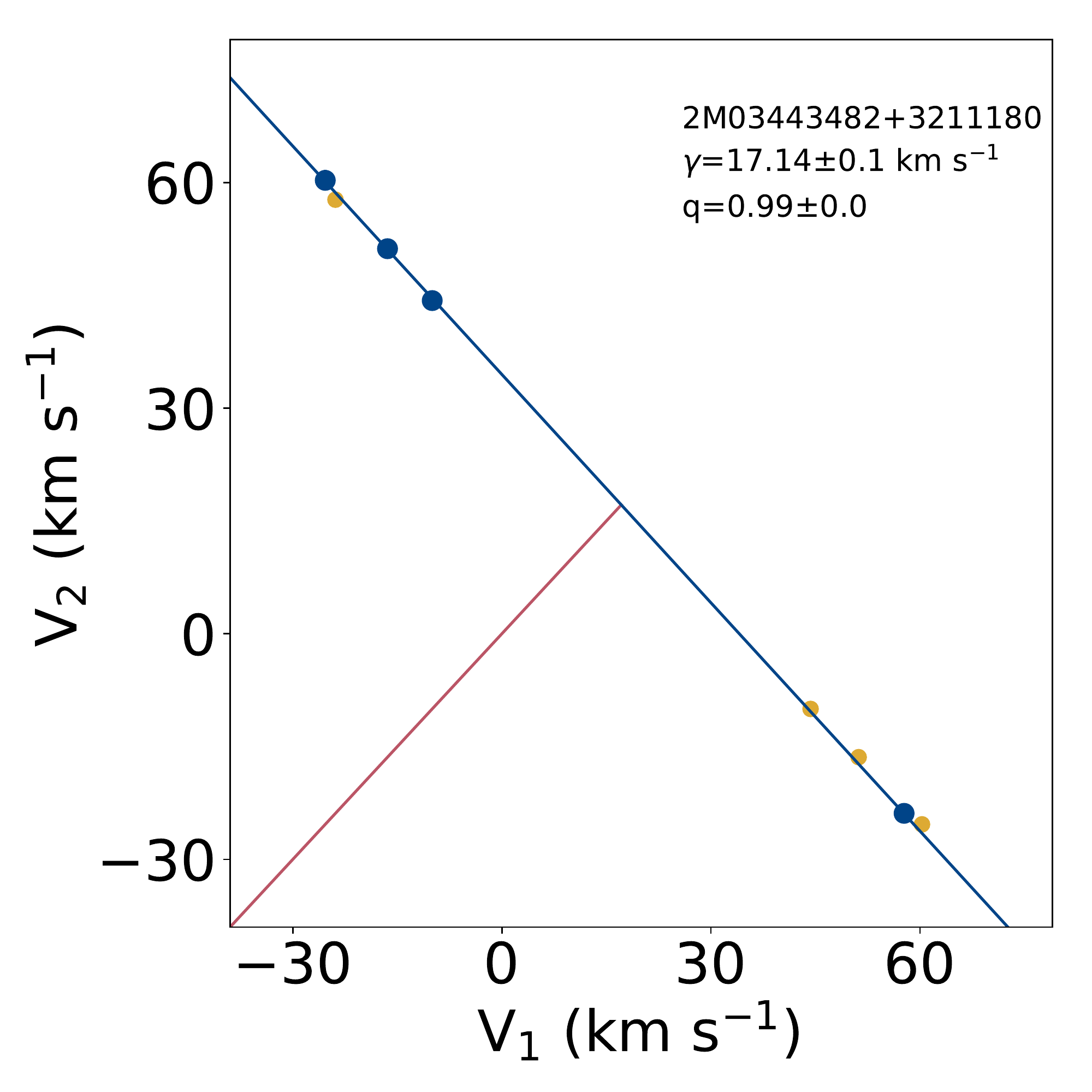}{0.33\textwidth}{}
 \fig{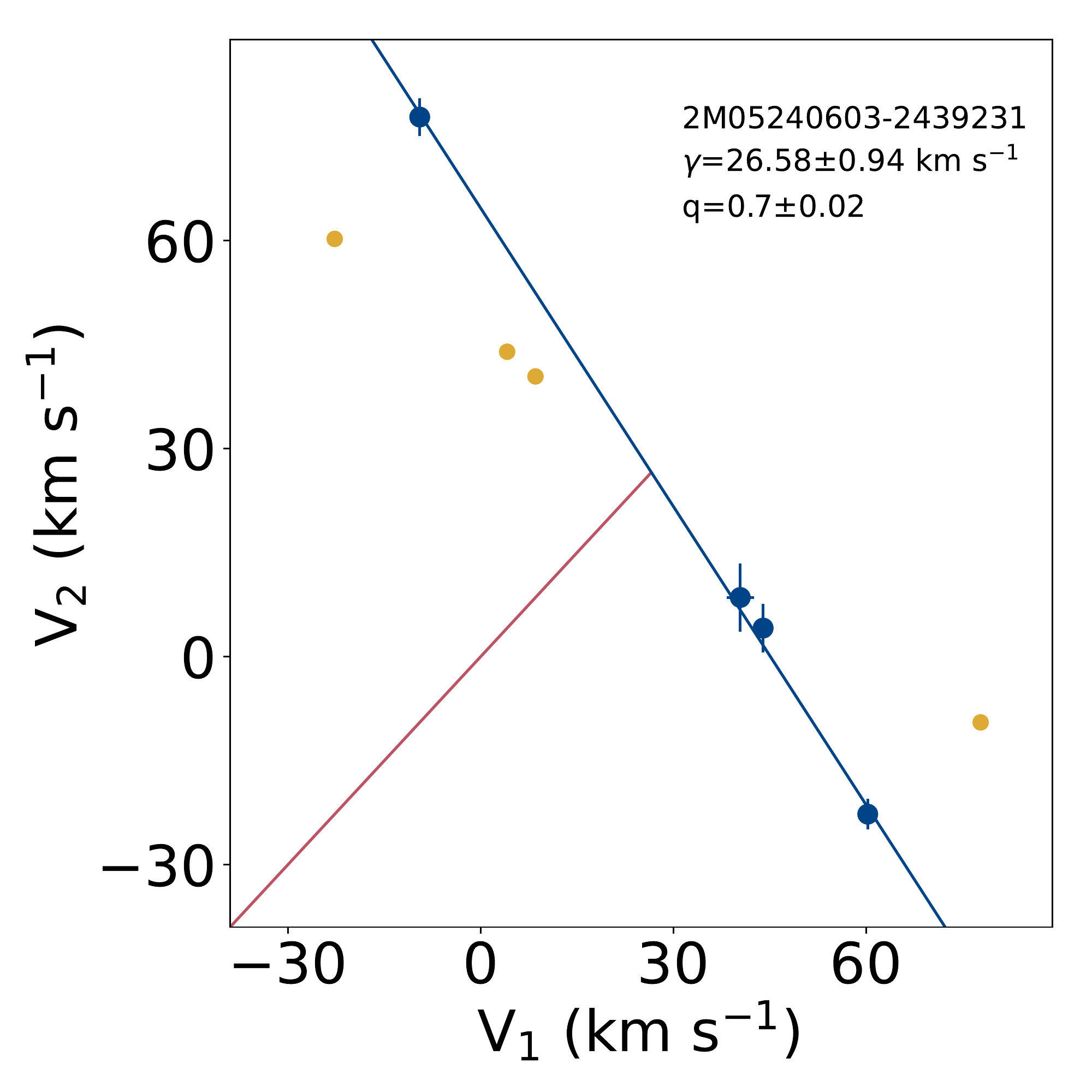}{0.33\textwidth}{}
 \fig{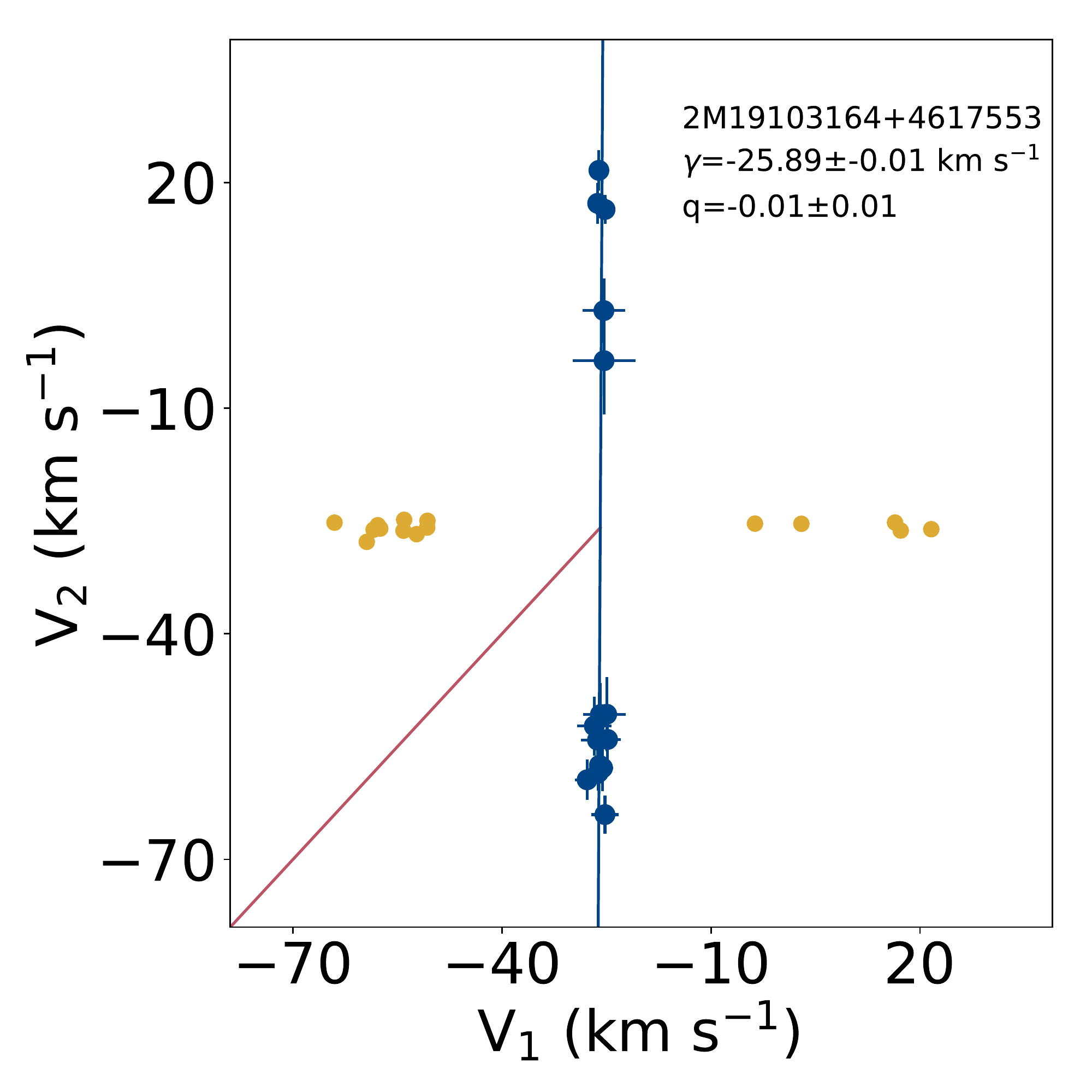}{0.33\textwidth}{}
 }
\caption{Example of Wilson plots. Blue dots show velocities of primary relative to the velocity of the secondary; yellow dots flip the assignment of the primary and the secondary for all epochs. Blue line shows the best fit to the data, the slope of this line relates to the mass ratio. Red line is the line of equality between $v_1$ and $v_2$, the intersection of these two lines corresponds to the barycenter velocity of the system. Left panel shows an example of an equal mass binary. Middle panel shows an example of system with lower $q$. Right panel shows an example of a system that likely contains a hidden tertiary companion. Note that in the equal mass systems there may be confusion in the assignment of a primary and a secondary, whereas in systems with lower $q$, an epoch with a wrong assignment can be identified through deviation from the best fit line.
\label{fig:wil}}
\end{figure*}

It is possible to determine the mass ratio ($q$) and the central velocity ($\gamma$) of most SB2s even in cases when there is an insufficient number of epochs to construct a full orbital fit. One of the ways to do it is to construct a Wilson plot \citep{wilson1941}, i.e., plotting velocities of the primary vs the velocities of the secondary (Figure \ref{fig:wil}). The slope of the linear regression fit to all these datapoints is equal to $-1/q$. Similarly, $\gamma$ is found as the point along that fitted line where $v_1=v_2$.

Although, generally, the primary and secondary of each system can be automatically identified based on the heights of their CCF profiles, some of the epochs in some of the systems may end up mislabelled. In systems with $q\lesssim 0.9$, Wilson plots can be used to diagnose whether it may be more appropriate to switch the assignments of the two components at any given epoch, to ensure a self-consistent slope. Similarly, this helps in assigning an RV to either a primary or a secondary (or both) in cases where only a single component is detected in the CCF, due to a combination of a low flux ratio, low signal-to-noise spectrum, and/or low velocity separation at a given epoch. 

The Wilson plot cannot help to confirm the RV assignment to the primary and secondary components of systems with $q\sim1$, however, as switching the assignment of the primary and the secondary would not significantly change the relative position of the data points with respect to the linear fit to them. Nonetheless, $q$ remains relatively well constrained in such systems, regardless of the confidence of the velocities assigned to each component.

We visually examined Wilson plots for all of the identified SB2s to identify systems in which $q$ can be confidently measured, and to the best effort corrected the assignment of the primary and the secondary in cases it was apparent they were mislabeled. This was not done with SB3s and SB4s due to the greater complexity of such systems.

Most of the systems in this sample have $q\sim1$, as expected from prior studies of binary populations \citep[e.g., ][]{moe2017}, as well as detection biases, where a significant imbalance in the flux ratio of two stars will make a companion more difficult to detect as an SB2. However, a suprisingly sizable number of systems (278, or $\sim5$\% of all the systems for which mass ratio has been measured) have inferred $q<0.1$, or even appear to have negative $q$, which would be physically impossible. Such systems are likely higher order multiples that have an unseen third (or fourth) companion. While we include the resulting fit in Table \ref{tab:system} for completeness, in such cases the assumptions made by the construction of the Wilson plot no longer apply.

\begin{figure}
\epsscale{1.2}
\plotone{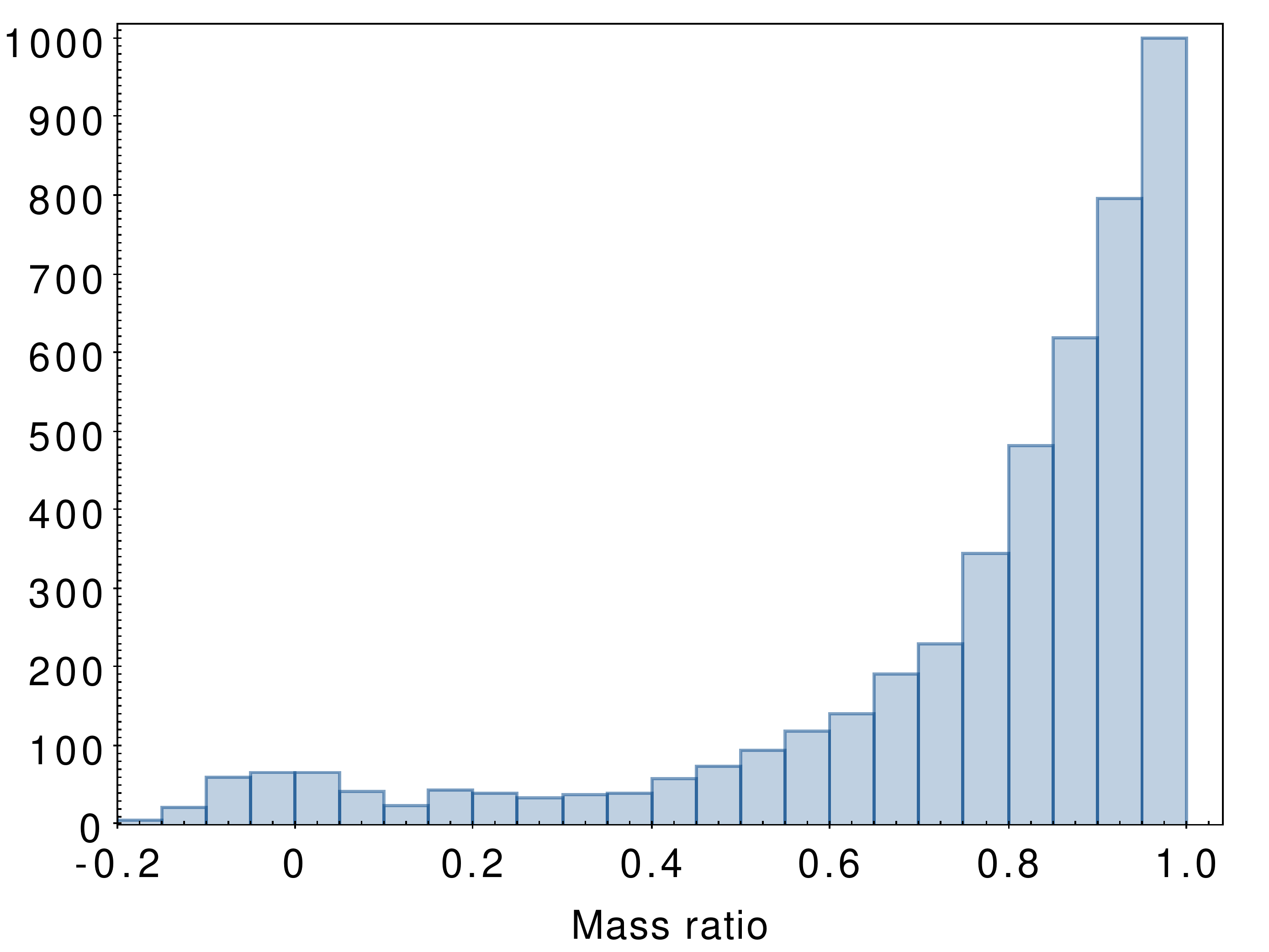}
\caption{Distribution of mass ratios in the sample, measured using Wilson plots. Sources with $q>1$ have been inverted as $1/q$ to be included in this plot. Sources with $q<0.1$ likely have an unresolved tertiary companion.
\label{fig:q}}
\end{figure}

We also note that, in some cases, it is possible to have fitted $q>1$. This occurs when the less massive component has a systematically larger height in the CCF profile across all epochs. For main sequence binaries, this would not typically occur: in these systems, the mass ratio and flux ratio should be correlated, and thus the source producing a larger CCF peak can be safely assessed to be the more massive component as well. Prominence in the CCF is still only serving as a proxy for mass, however, and this proxy status can and does fail. This may occur for instrumental reasons, e.g., if the cross-correlation template has a better match to the secondary rather than the primary, resulting in a stronger peak for the fainter and less massive component. Alternatively, there may be an astrophysical explanation for the inversion of the flux and mass ratios in more exotic systems, such as 2M17091769+3127589 (Miller. A, in prep). This system was successfully deconvolved as an SB2 in APOGEE DR14 data, but, unfortunately, not in subsequent data releases (and thus it is excluded from the catalog presented in this work). This system contains a post mass transfer red giant, which dominates the spectrum at infrared wavelengths and is thus assessed to be the primary component in our flux-based analysis. The evolved component has lost much of its mass, however, to its less evolved companion. This less evolved source is now more massive, but still fainter at infrared wavelengths, and is thus identified as the secondary in our flux-based assignment, producing $q>5$. To better highlight other such systems that may be present in the dataset, we preserve the systematic assignment of labels of $v_1$ and $v_2$ (other than ensuring the aforementioned self-consistency between epochs), and thus allow $q$ to extend $>1$ in Table \ref{tab:system}. We also present the mass ratio distribution graphically in Figure \ref{fig:q}; to avoid distorting the x-axis to include a small number of q $>$ 1 sources we remap those mass ratios into the 0--1 range by presenting them as 1/$q$.

\subsection{Orbital fitting}

To estimate orbital parameters for the identified systems, we have used a branch of the The Joker \citep{price-whelan2017} adapted for SB2s. The Joker performs Monte Carlo sampling over a range of possible parameters in order to identify a series of likely orbits. It has previously been used on the APOGEE data to perform statistical analysis over the orbital posterior distribution on sparse RV curves that may not necessarily be characterized by a single period, as well as fitting orbits of a number of SB1s \citep{price-whelan2018,price-whelan2020}.

SB2s are more complex than SB1s. As discussed in the prior section, it is not always straightforward to definitively assign measured velocities to an SB2's primary or secondary component. Furthermore, in some cases where two components have a relatively small RV difference even at maximum separation, rather than measuring the velocities of individual stars, or even the velocity of a primary, the measurement is pulled closer to the barycenter of the system. Both of these effects can confuse the orbital fit, even if a system has been observed over 20+ distinct epochs. On the other hand, as the orbital equation for SB2s consists of 7 parameters, and each epoch can provide two separate data points, in principle, it may be possible to unequivocally solve a complete orbit with as few as 4 optimally timed epochs (though, such cases are rare). 

To derive the range of orbital solutions that may be appropriate for each system, we initialized The Joker with 5 million samples for all 1490 SB2s with at least 4 epochs and a well fit mass ratio (Section \ref{sec:q}).

For some systems, if the priors are well-constrained or if the RV uncertainties are large, it is possible to use The Joker to estimate uncertainties in the orbital parameters. However, it is not always practical or computationally expedient. The orbits of many SB2s are so tightly constrained that, regardless of the initial number of samples, The Joker would not return a family of similar orbits with scatter in their parameters, but rather a singular orbit, even with an initial samples of a few millions for each system.

Thus, to estimate formal errors, we further fit the identified SB2s with the IDL code \texttt{rvfit} \citep{rvfit}. Unlike The Joker, which can sample a wide range of orbital parameter space, \texttt{rvfit} is a fitter, and as such, it may struggle to find appropriate parameters on its own, getting stuck in local minima. However, given appropriate initial estimates, such as those returned by The Joker, it can not only improve the fit but also estimate uncertainties in a more robust manner, through the direct fit and a subsequent MCMC sampling that automatically sets appropriate priors in the parameter space from the fit.

\begin{figure}
\epsscale{1.2}
\plotone{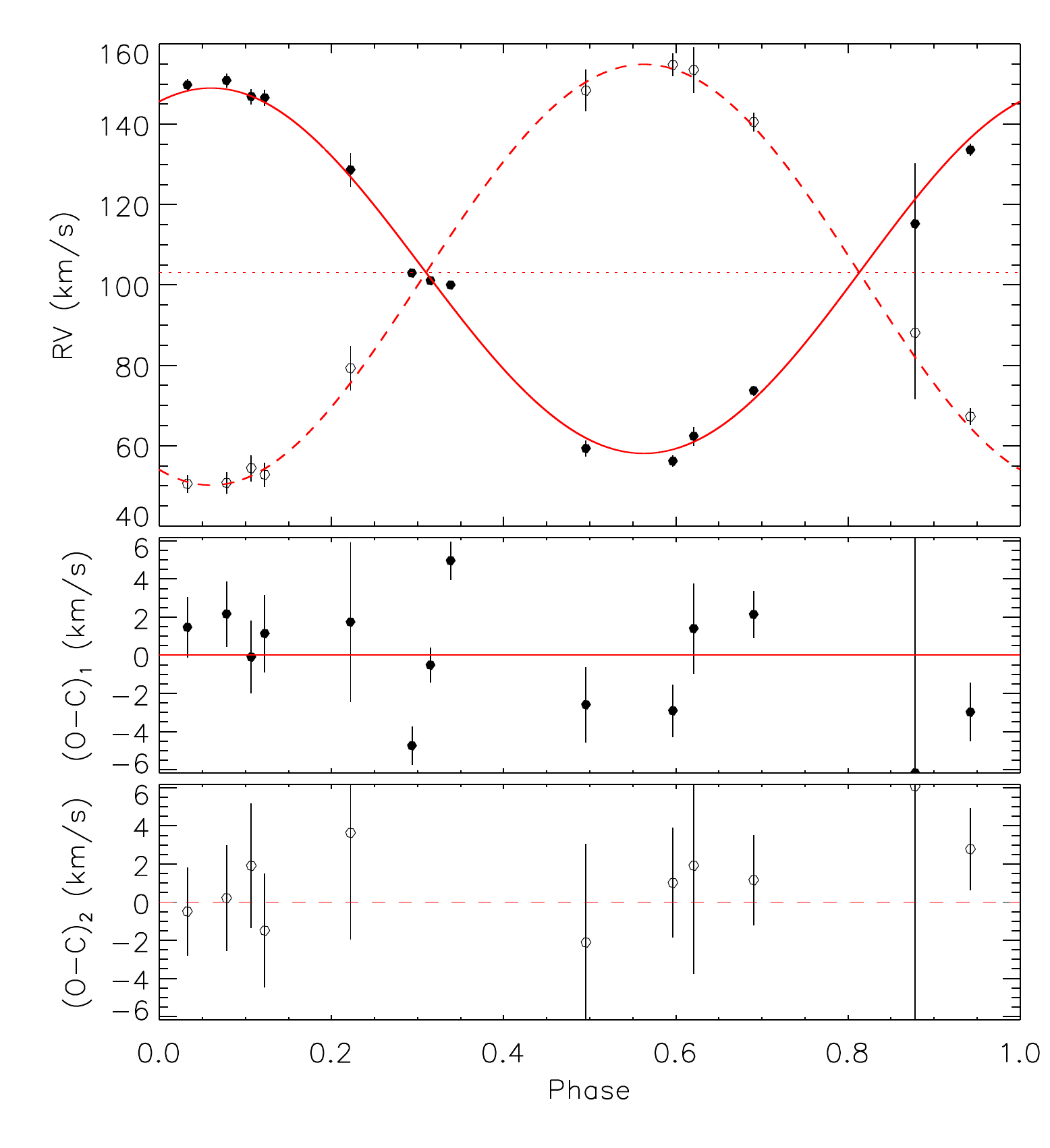}
\caption{Orbital fit for 2M04442280-7524067, shown as an example of a typical system with a well-constrained solution. The top panel shows the orbit, the bottom two panels show the residuals of the fit for the primary and for the secondary.
\label{fig:orbit}}
\end{figure}

Through the combined usage of \texttt{rvfit} and The Joker, we visually examined the fitted orbits and the resulting parameters of all SB2s with at least 4 epochs to ensure they could be well fit by only a single mode of orbital solutions. This resulted in a sample of 325 well fitted systems (an example of such a system is shown in Figure \ref{fig:orbit}).

We note that, outside of APOGEE, there are a number of other surveys and smaller studies that may have obtained spectra and measured RVs for these systems. In some of these systems, these additional RV measurements may be sufficient to fully solve the orbit. For the sake of having a uniform catalog, these additional data are not included in the orbital fitting presented in this work. In the future, however, performing a cross-match with other data \citep[e.g.,][]{pourbaix2004} should make it possible to further increase the census of SB2s with full orbital solutions.

\subsection{Light curves}
\begin{figure}
\epsscale{1.2}
\plotone{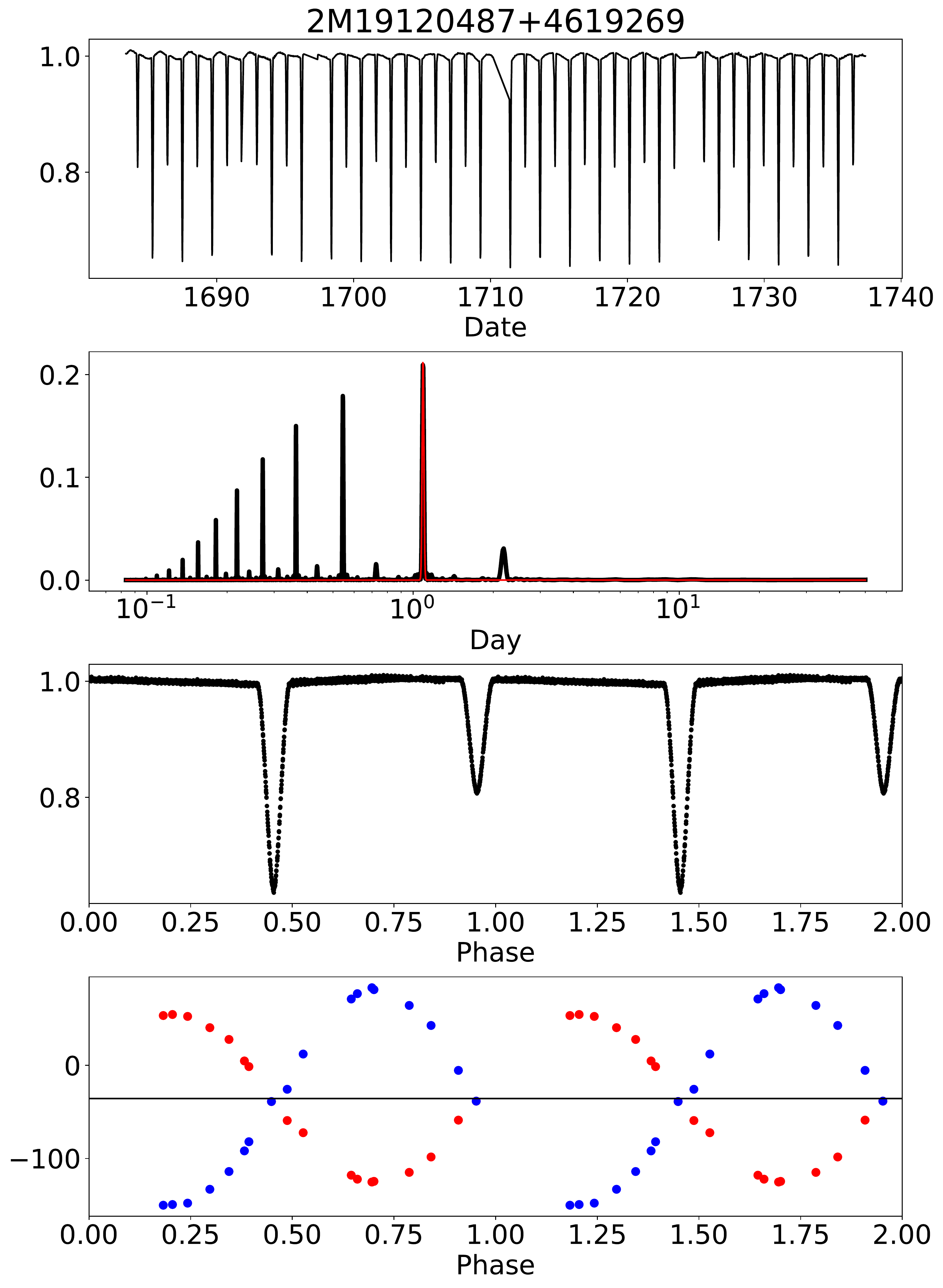}
\caption{Example of a spectroscopic binary that is also eclipsing. Top panel shows the full TESS light curve. Second panel shows the Lomb-Scargle periodogram, in this case the strongest periodic signal originates from a half of the full period of the system. Third panel shows the phase-folded light curve, and the bottom panel shows the phase folded RV curve.
\label{fig:eb}}
\end{figure}
Some of these systems, in addition to their spectroscopic binary signatures, also exhibit eclipses. To identify this subset of the sample, we used the eleanor pipeline \citep{feinstein2019} to extract light curves for all systems observed by TESS to date. These light curves were produced from the full frame images spanning the first 26 sectors of the survey. In total, 5485 sources fall into the TESS footprint, the remaining sources have not yet been observed.

We then examined all of these light curves to see if they exhibited a periodic signature. Regions of the light curve that had bad systematic trends that were not well corrected by eleanor, or regions that contained significant artifacts not properly flagged were manually removed using a custom Python code. Following this, we calculated a Lomb-Scargle periodogram \citep{lomb1976,scargle1982} for each light curve to find a dominant period. Sometimes, this dominant period is a factor of 2 (or 4) shorter than binary's true orbital period - in these cases, we visually corrected period to produce a clean folded light curve. On occasion, significant out of eclipse variability prevented the Lomb-Scargle periodogram from returning the correct period, despite clear eclipses present in the data. In these cases, we measured the period using two neighboring eclipses, ensuring that this period would correctly fold other observed eclipses as well.

Close eclipsing binaries (such as contact, semi-detatched systems, or systems with strong ellipsoidal variations) can be easy to identify in the light curves. Contact binaries, on the other hand, may have similar signatures to rotating spotted stars. Furthermore, even if the system is not truly eclipsing, tidal synchronization of the two stars may force the rotational to be the same as the orbital period in some cases. For this reason, we have flagged all light curves that appear to be periodic.

\begin{figure}
\epsscale{1.1}
\plotone{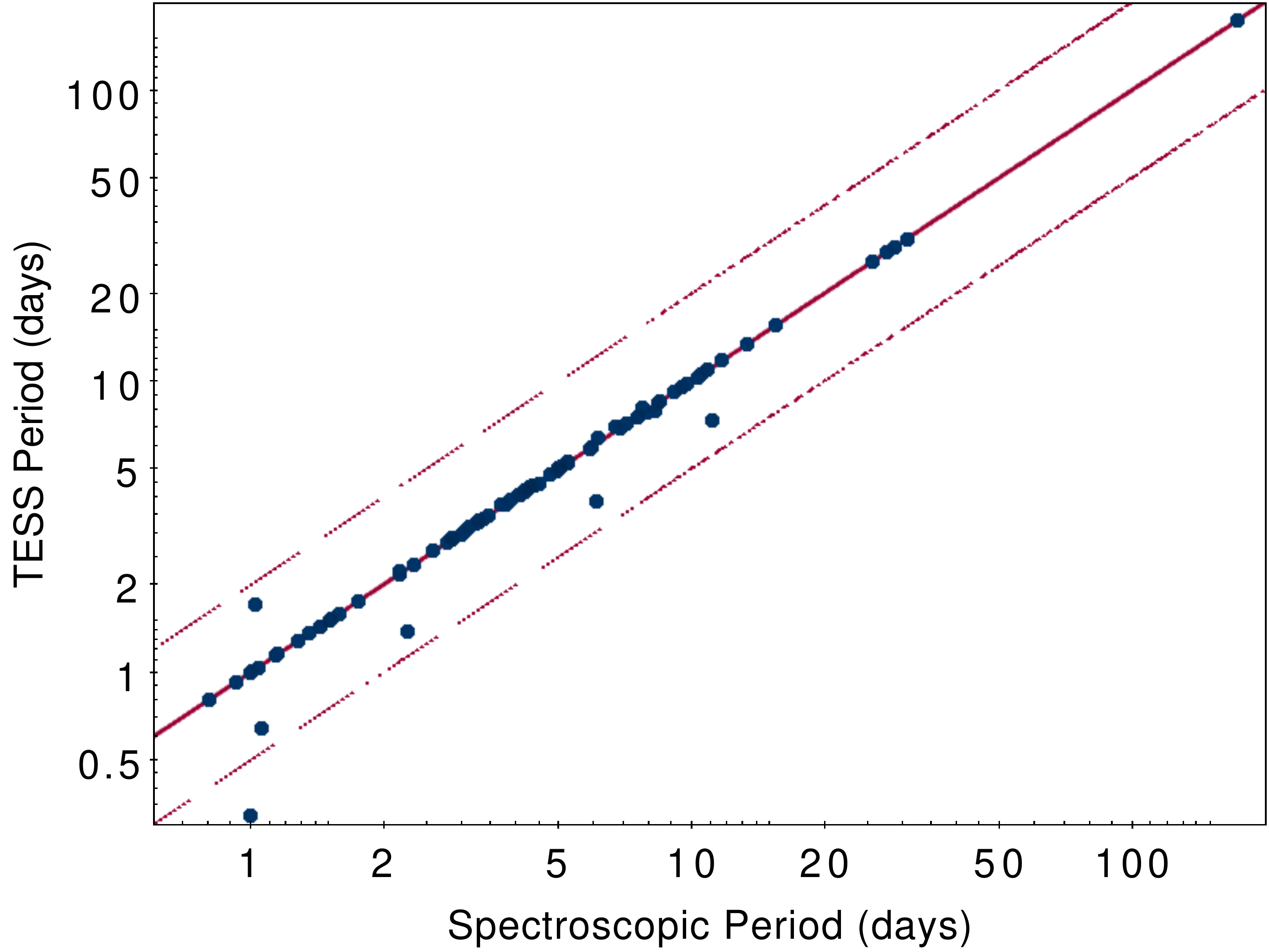}
\caption{A comparison of the period derived from the orbital fit for the SB2s to the photometric period derived for the eclipsing binaries in TESS data. The solid line shows one-to-one relation, dashed lines show x2 and x0.5 offsets. The outliers do not have light curves consistent with being detached eclipsing binaries and are more likely to be rotating variables. The measurements that are consistent between two methods agree in the period determination to within 0.5\%.
\label{fig:pp}}
\end{figure}

In total, out of 5485 sources with TESS light curves, 1504 have periodic light curves. Most of these sources are not necessarily eclipsing, but rather have periodic signal from other activity, such as rotation. They are included in the catalog because it can be difficult to separate contact binaries from rotators. However, 369 have clear signatures of being detached eclipsing binaries. Of the 320 systems with fully fitted spectroscopic orbits, 88 are periodic; of those, 82 have a good match between the photometric and spectroscopic periods. The remaining 6 are not classified as detached eclipsing binaries. Their photometric periods do not relate to the spectroscopic orbit: rather, the periodic signal may be related to the rotational period of individual stars, may be affected by poor signal-to-noise, or may be contaminated by a presence of neighbors in the TESS aperture (Figure \ref{fig:pp}).

There are a number of other surveys of variable stars that may provide additional independent constraints on these systems' orbital periods. For example, the Variable Star Index \citep[VSX][]{watson2006} currently includes more than 2 million variables, of which 868 coincide with spectroscopic binaries in our sample. Only 323 of these are also identified as periodic variables in TESS: the remaining sources are either outside of the TESS footprint or have preferentially longer periods. The sources we identified as detached eclipsing binaries via TESS light curves are preferentially identified as EAs in VSX. There is also a good agreement in the derived periods, although outside of the sources for which full orbits have been constructed with the help of RVs, periods measured with TESS do appear susceptible to a multiplicative factor of 2 offset.

The systems that are eclipsing SB2s offer a good opportunity to better constrain the stellar mass-radius relationship, as orbital fit of such systems is the most direct method of measuring these parameters \citep[and references therein]{serenelli2021}. Currently, the systems for which we have fitted full spectroscopic orbits only need their light curve characterized with tools such as Phoebe \citep{conroy2020} or ellc \citep{maxted2016}. The sources that have been flagged as detached eclipsing binaries but have an insufficient number of spectroscopic measurements to fit an orbit would benefit from follow up observations, particularly among some of the sources that are pre- and post main sequence stars. Furthermore, although this was not done in this work, it is possible to force the eclipsing period to the spectroscopic orbit, which may make it possible to derive masses with a minimal number of RV measurements.

\begin{figure}
\epsscale{1.1}
\plotone{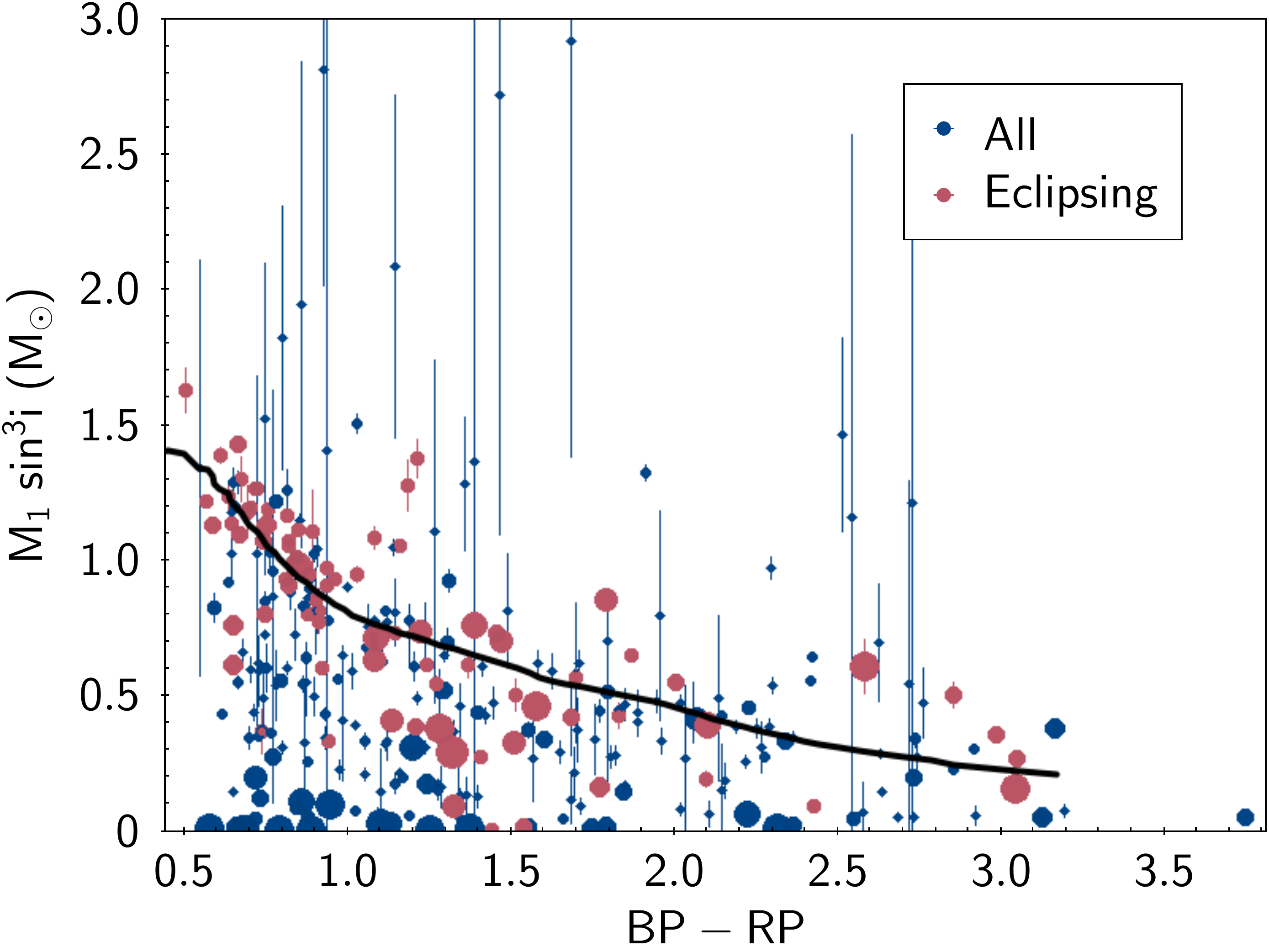}
\plotone{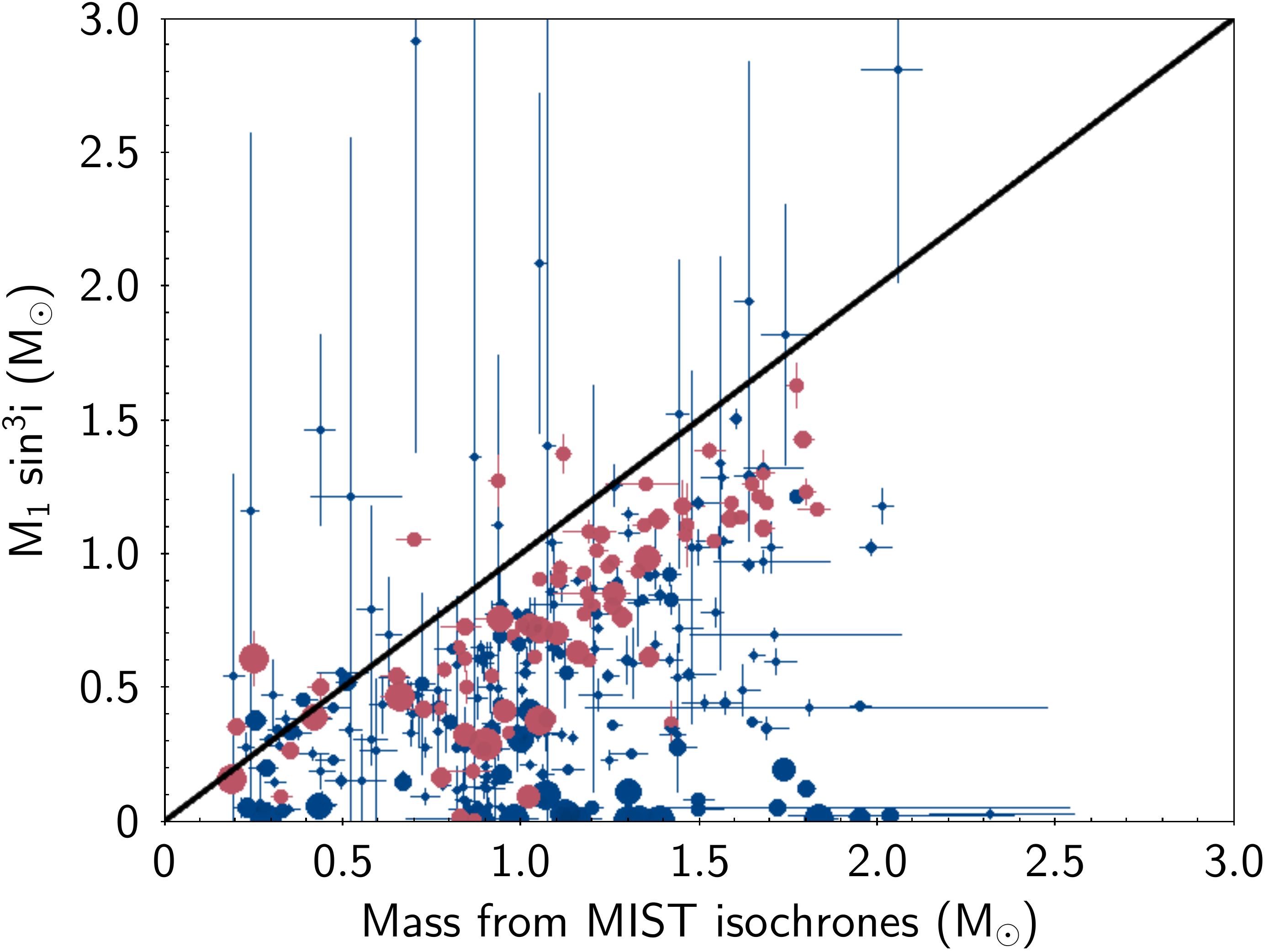}
\plotone{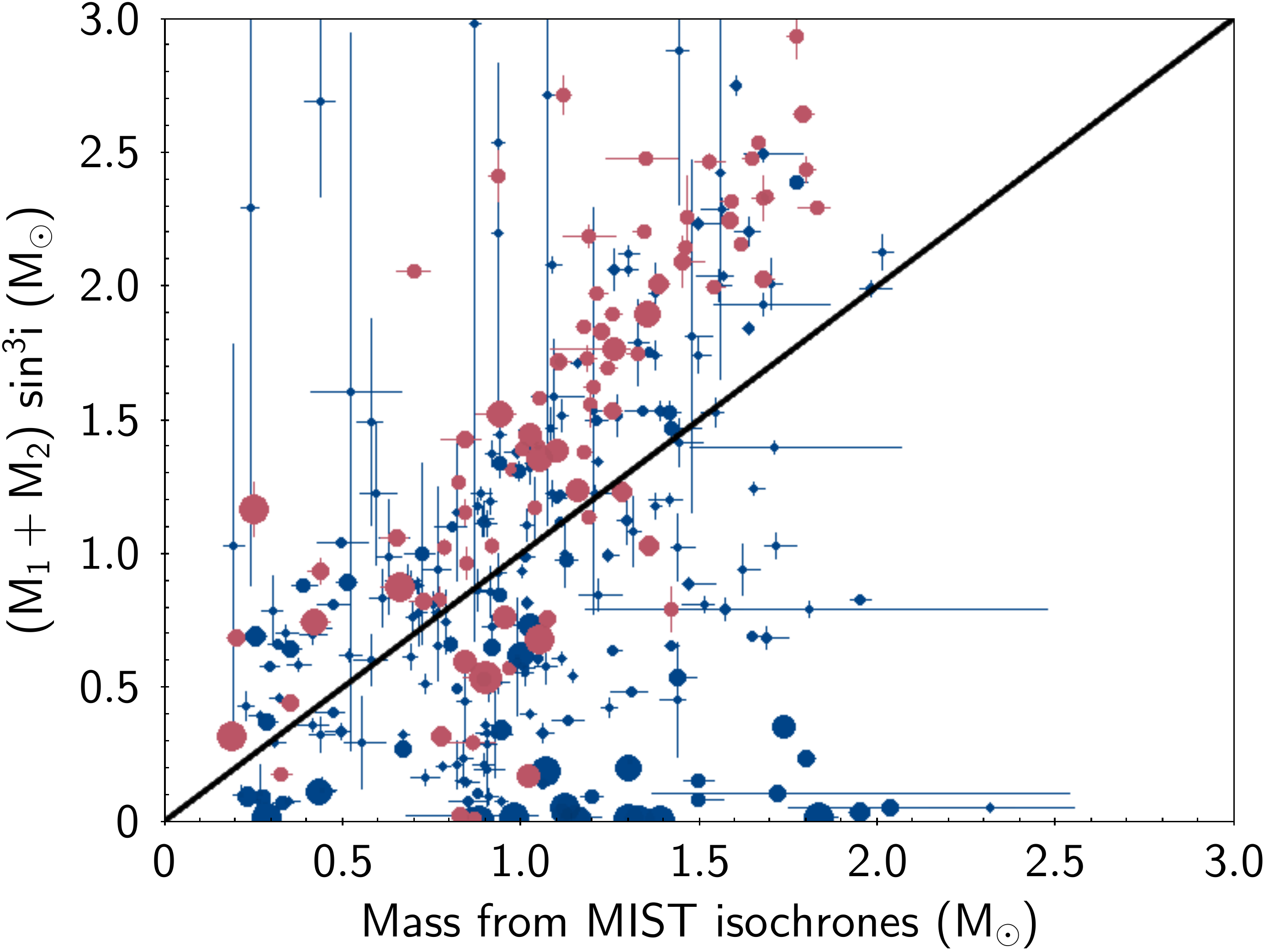}
\caption{Top: Distribution of M$\sin^3 i$ of the primaries derived from the orbital fitting, as a function of color in Gaia bandpasses. The black line shows the mass-color relation for the main sequence stars from MIST isochrones \citep{choi2016}. Middle: derived M$\sin^3 i$ of the primary versus the inferred mass from isochrones . Bottom: combined M$\sin^3 i$ of the primary and the secondary versus the inferred mast from isochrones. Systems that have been identified as eclipsing (and should have $i\longrightarrow90^\circ$) are highlighted in red. The size of the symbols is inversely related to the orbital period, with the larger circles having shorter period.
\label{fig:cm}}
\end{figure}

Even without performing a detailed light curve analysis, eclipsing binaries can strengthen the mass estimates inferred for SB2s. The RV orbital fit does not directly solve for mass; rather, it provides $M\sin^3 i$ for all the components, such that the system's inclination must be determined to uniquely identify the components' masses. Since EBs are seen close to edge on, with $i\longrightarrow90^\circ$ (though eclipsing configurations with $i$ as low as $60^\circ$ can be possible), this ambiguity in the system's inclination is significantly reduced, such that the inferred $M\sin^3 i$ value approaches the true $M$. This can be seen in Figure \ref{fig:cm}, in examining color versus $M\sin^3 i$, which is, on average, higher for eclipsing binaries at a given color, forming an upper limit to the relation for the main sequence dwarfs. A few sources may have $M\sin^3 i$ values above this limit, either due to very uncertain measurements or because the source has evolved off the main sequence. Converting colors to masses using isochrones does not perfectly reproduce the $M\sin^3 i$ relation measured from the orbital fits, which are systematically offset from the line of equality. This may be because the stars largely inhabit the binary sequence, and thus their colors do not serve as an accurate proxy for the mass of a single star. On the other hand, the mass derived from the isochrones underestimates the total mass of both stars.

\section{Discussion}\label{sec:discussion}

\subsection{Comparison to other catalogs}

Over the history of the APOGEE survey, there have been a number of efforts dedicated to the search for SB2s. They include works such as \citet{fernandez2017} and \citet{kounkel2019} for the pre-main sequence stars, \citet{skinner2018} for the M dwarfs, and \citet{mazzola2020} - presenting SB2 candidates in the APOGEE DR14 sample as a preliminary version of the catalog in this paper. One of the most extensive previous efforts, however, was never published and only available as an online database\footnote{\url{http://astronomy.nmsu.edu/drewski/apogee-sb2/apSB2.html}}. This includes 1,208 candidates identified through visual inspection of all of the CCFs of the APOGEE-1 survey of the data obtained through DR12. Our automated code flags 942 of these systems, and 1544 systems in total, with the sample restricted to what has been observed up to July 2014 (which was the cut-off date for DR12). We note that CCFs were computed differently for DR12 and the current internal data release, even for the same system, and some of the constructed CCFs can be more or less sensitive to the presence of companion, depending on the parameters used to select the best-matched template spectrum for which the CCF was computed.

Another notable work identifying SB2s in the APOGEE data was done by \citet{el-badry2018}. However, instead of analyzing CCFs, they performed multi-template fits, identifying systems that are often unresolved in CCF space. As such, out of 3308 systems that they identify as multiple, we can only independently verify 702. The remaining $\sim$2600 are presumably binaries on wide orbits, whose velocity separation is too small for us to detect two independent CCF peaks. On the other hand, we identify only 42 SB2s among the 16,833 sources they flag as single stars, indicating their criteria for discriminating between single and multiple systems appears to be quite robust. The remaining systems we identify as multiple were not included in their analysis.

\citet{price-whelan2020} performed an analysis of radial velocity variable systems in the APOGEE data to identify 19,635 candidate SB1s, of which 1,032 comprise their ``gold sample'', i.e., systems with robust orbital solutions. In their full sample of candidate SB1s, there are 1154 that we find to be SB2s or higher order multiples. Of these, we measure orbits for 127, although only 30 of these coincide with the \citet{price-whelan2020} gold sample. In general, there is good agreement between their SB1 orbital solution and the solution we infer for the primary of the SB2 system, with the exception of 4 stars for which the APOGEE RVs may be corrupted due to their nature as SB2s.

There is some overlap between the sources targeted by APOGEE and those targeted by other surveys. Recently, \citet{traven2020} performed a search of SB2s in GALAH spectra. GALAH is an optical spectrograph, with $\sim$25\% higher resolution than APOGEE. They identified 19773 candidates out of 587,153 sources. It is difficult to compare the derived SB2 fraction between the two surveys, both due to the still proprietary nature of much of the underlying data, and due to the difference in the targeting strategy of the surveys (i.e., dwarfs make up a significantly higher fraction of the sources observed by GALAH in comparison to APOGEE). However, out of 539 sources observed by APOGEE and identified by \citet{traven2020} as SB2s, we can independently confirm 377 of them (of which 21 are SB3s). Similarly, of the 50 sources present in the publicly released GALAH DR2 \citep{buder2018} that we identify as multiples, \citet{traven2020} have also identified 30 of them.

\subsection{Relation to Gaia EDR3}

\begin{figure*}
\epsscale{1.1}
\plottwo{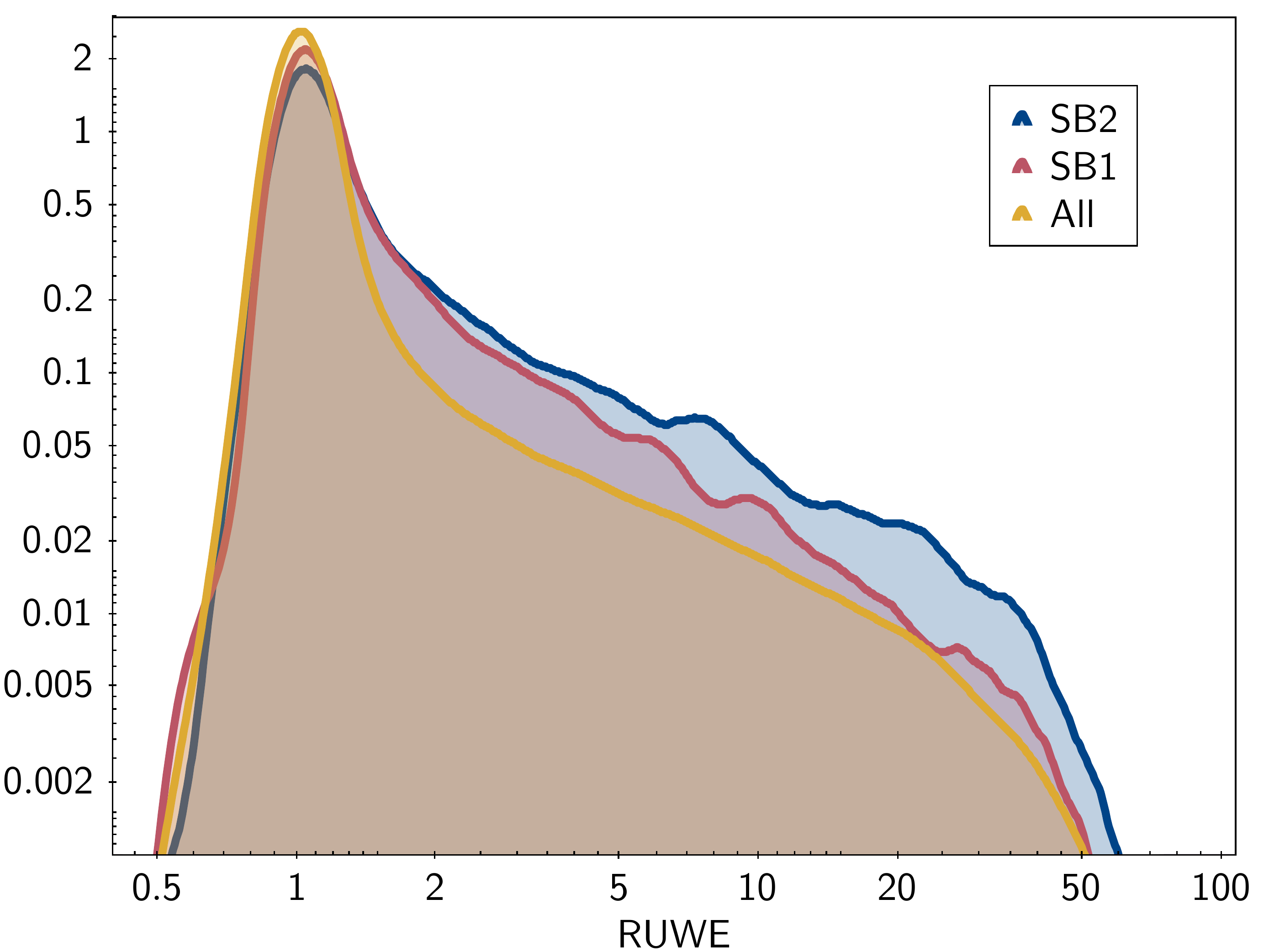}{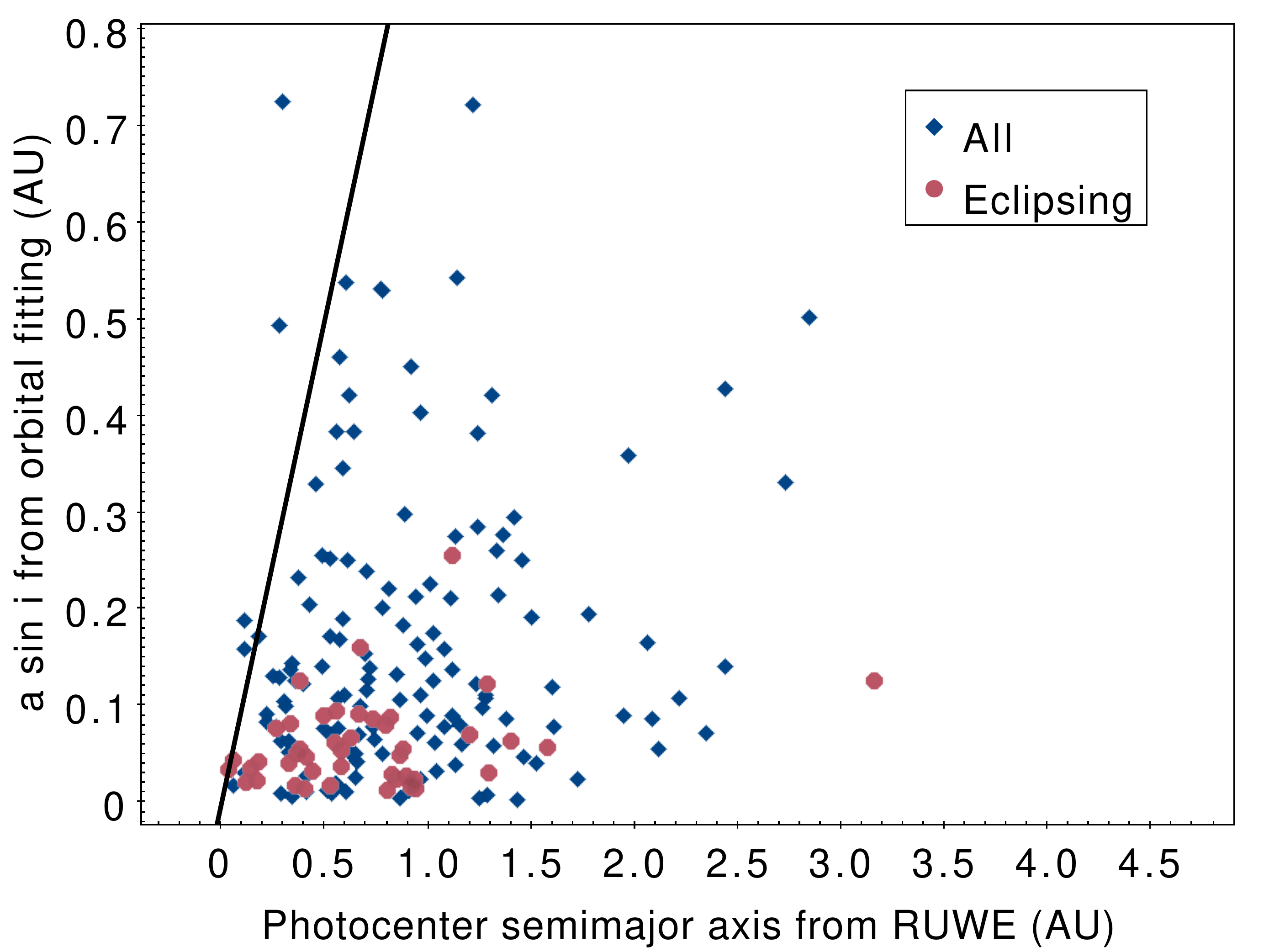}
\caption{Left: Kernel density estimation showing the distribution of RUWE from Gaia EDR3, with log axes in both dimensions. The full APOGEE sample (consisting in large part of single stars) is shown in yellow. SB1s/RV variable systems from \citep{price-whelan2020} are in red. SB2s (and higher order multiples) in this work are shown in blue. All three distributions are normalized by the area. Right: Semi-major axis $a\sin i$ derived from the orbital fitting of SB2s in this work versus photometric semi-major axis derived from RUWE using the expression from \citet{stassun2021} for sources with RUWE between 1 \& 1.4. The solid line shows one-to-one relation.
\label{fig:ruwe}}
\end{figure*}

\begin{figure}
\epsscale{1.1}
\plotone{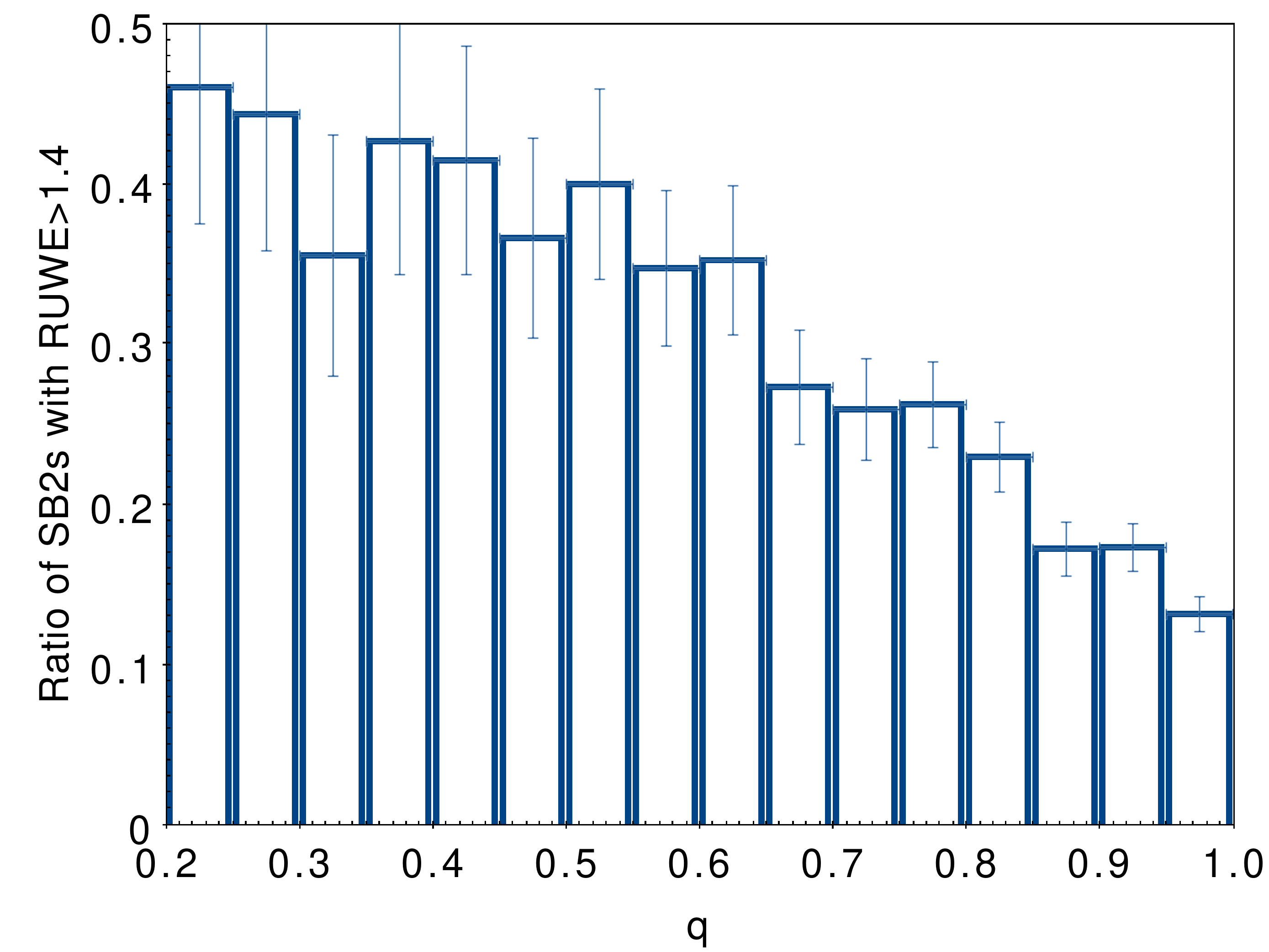}
\caption{Fraction of sources in the sample with high RUWE$>$1.4 as a function of mass ratio.
\label{fig:ruweq}}
\end{figure}

\begin{figure*}
\epsscale{1.1}
\gridline{\fig{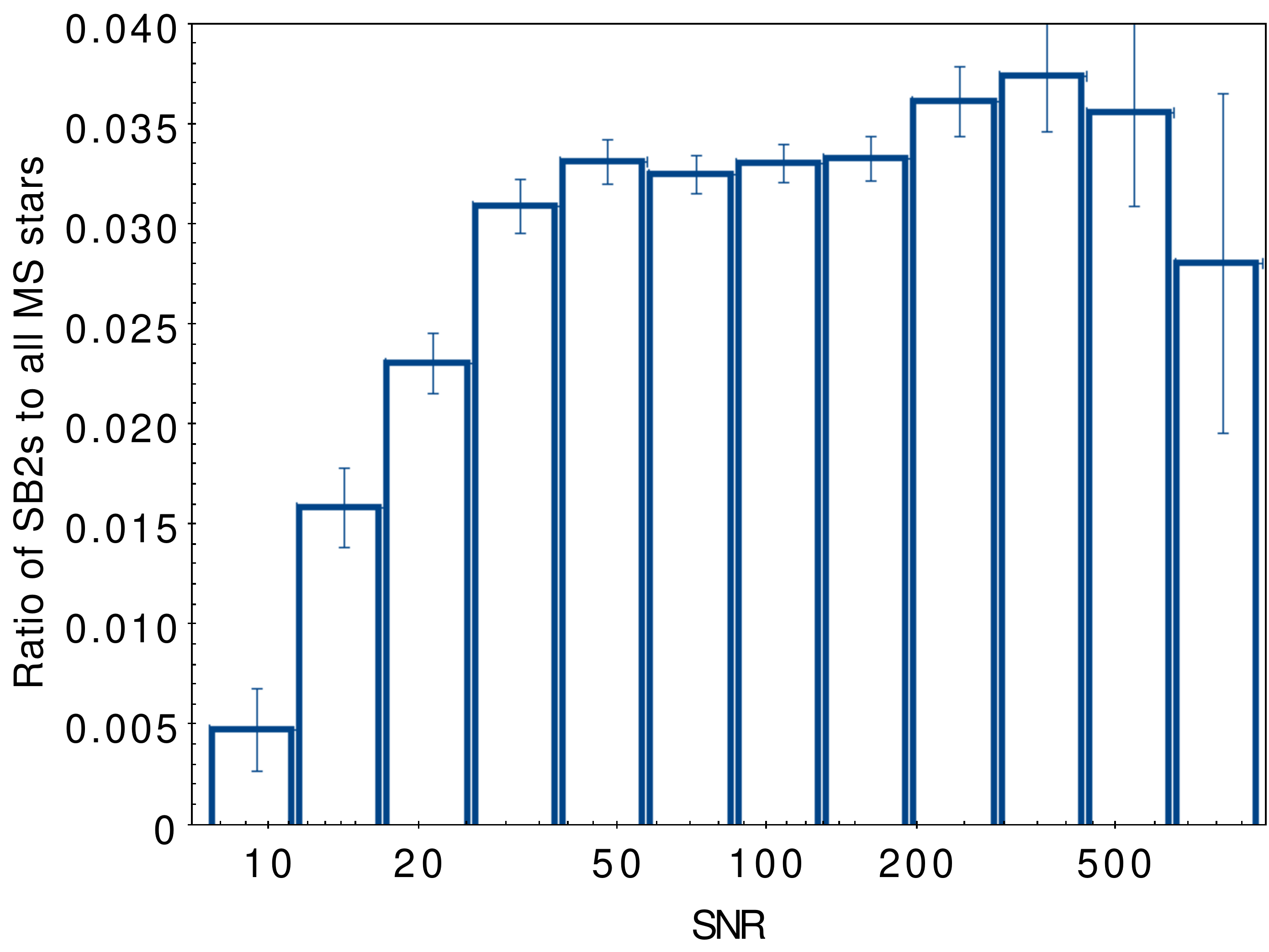}{0.33\textwidth}{}
 \fig{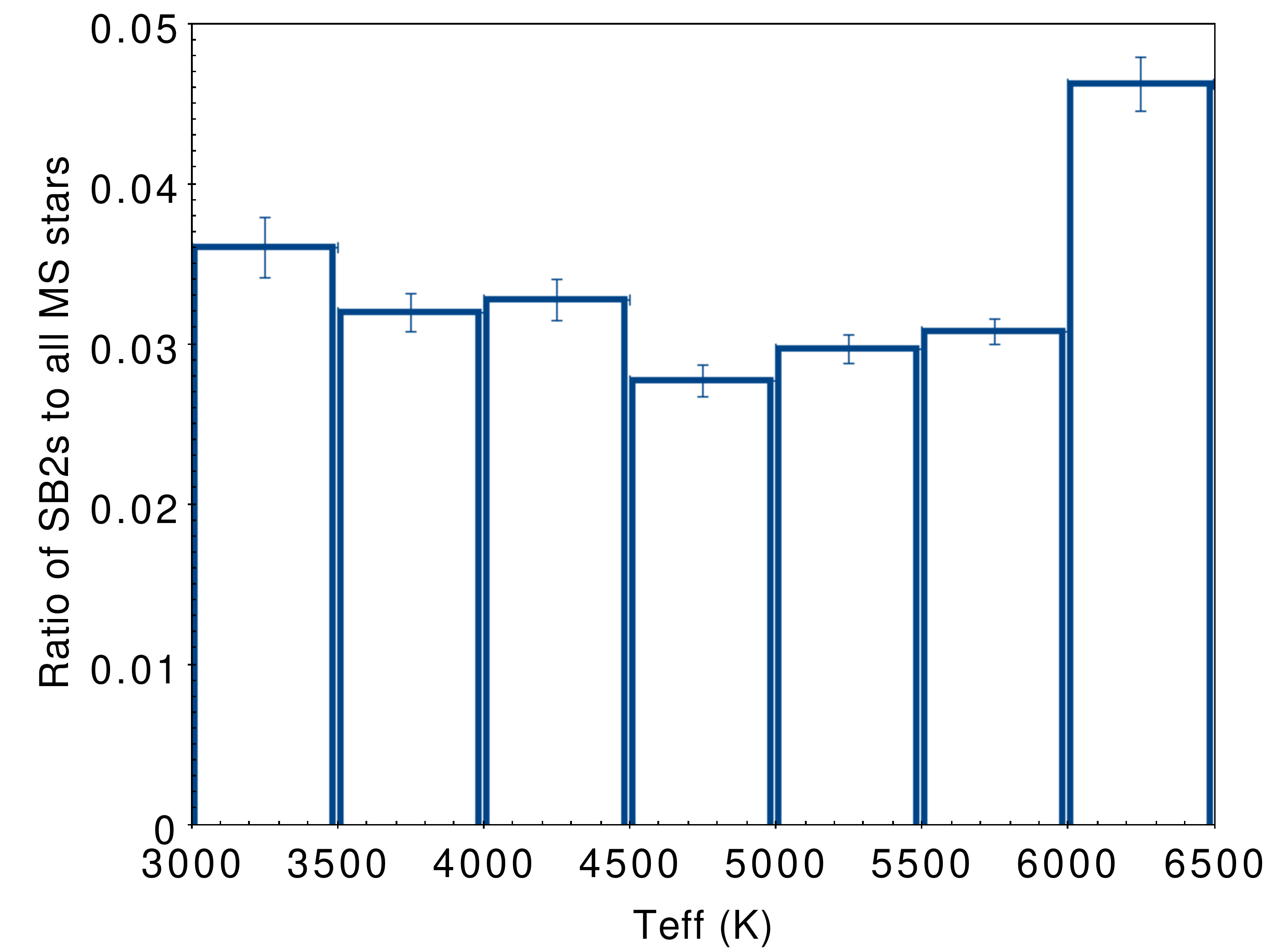}{0.33\textwidth}{}
 \fig{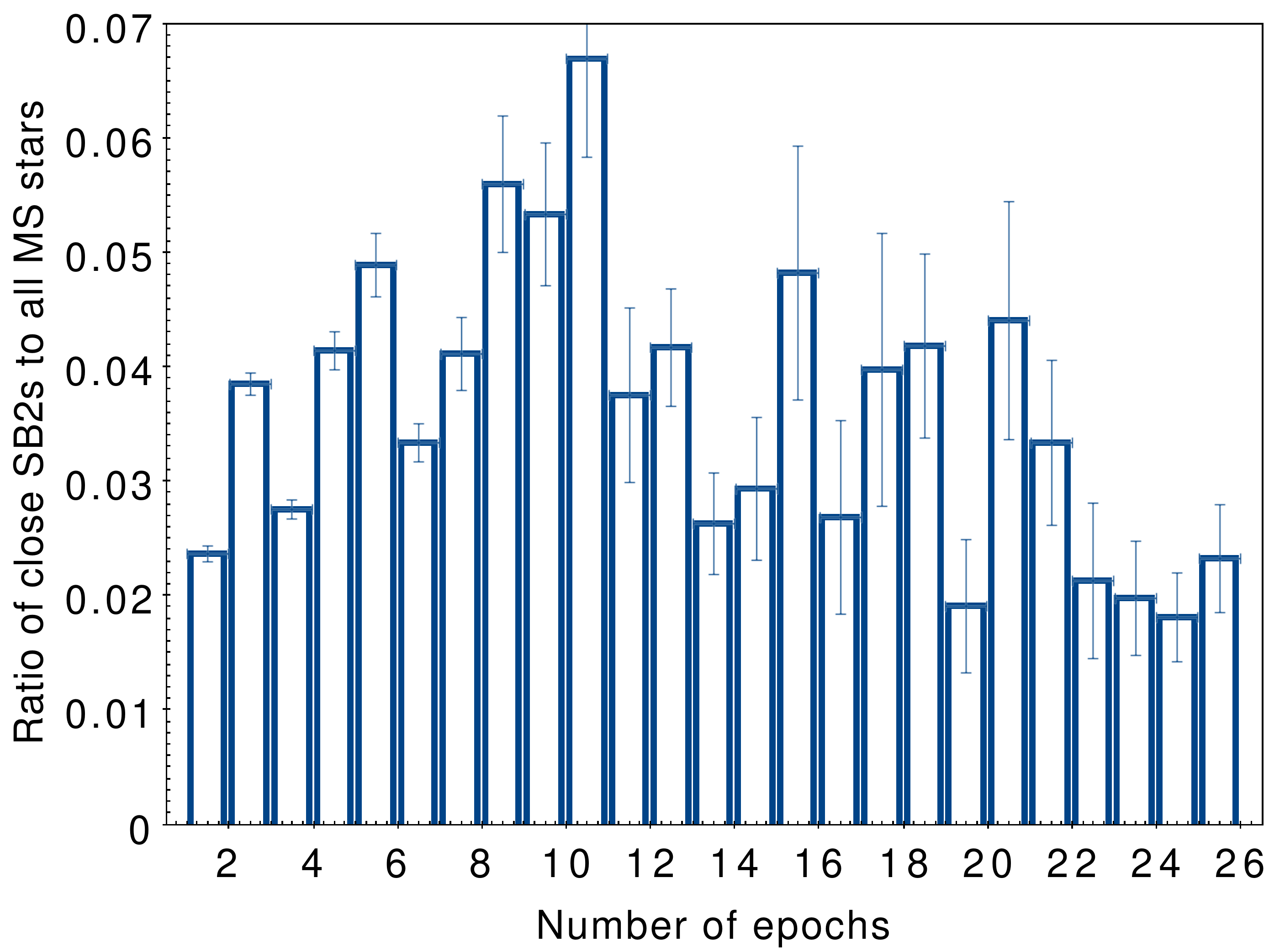}{0.33\textwidth}{}
 }
\caption{Fraction of sources identified as multiples in the full APOGEE sample of main sequence stars, defined by \logg$>4$ dex and \teff$<$6500 K, as a function of SNR, \teff, and the number of epochs.
\label{fig:ratio}}
\end{figure*}

\begin{figure*}
\epsscale{1.1}
\plottwo{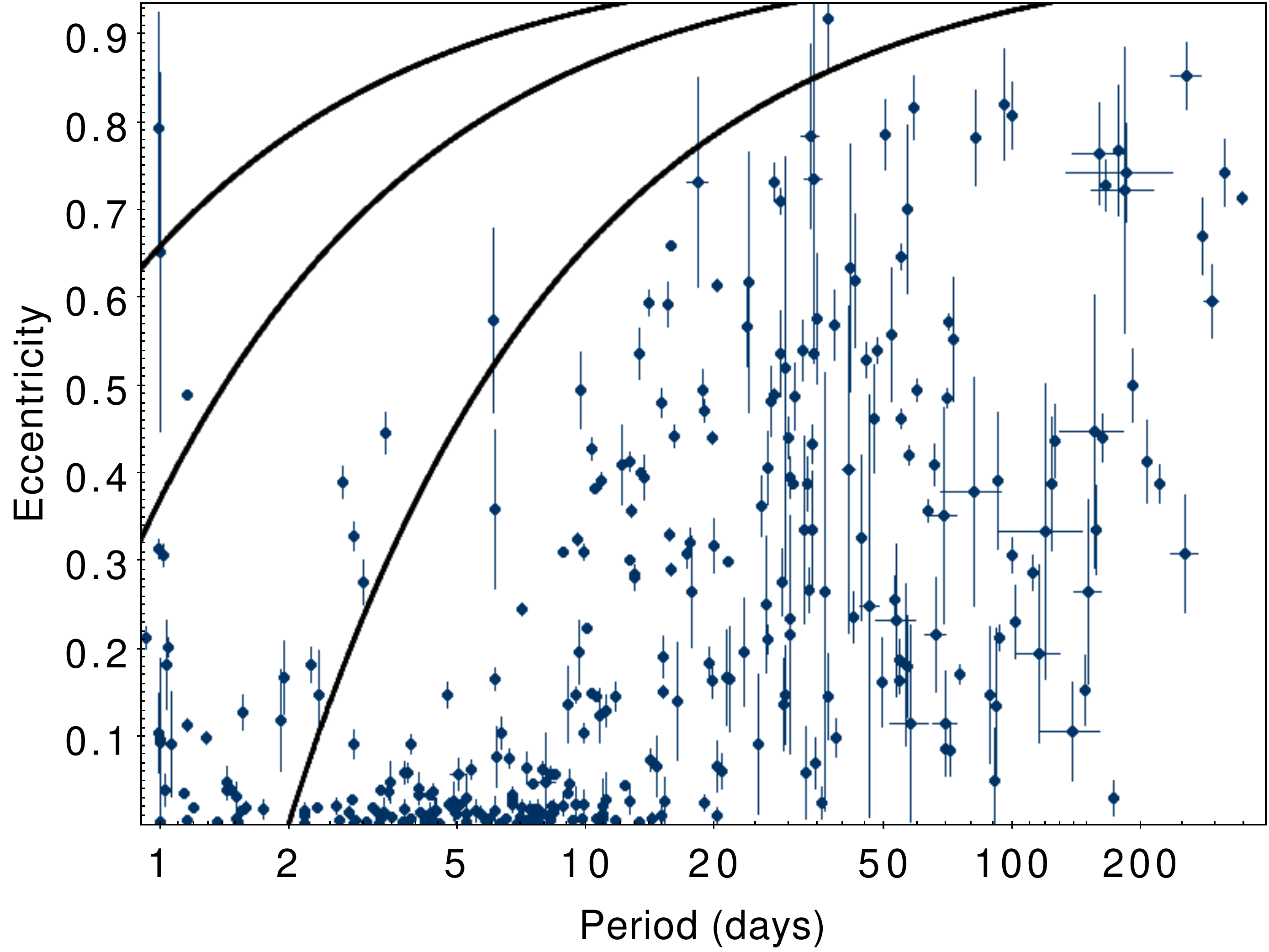}{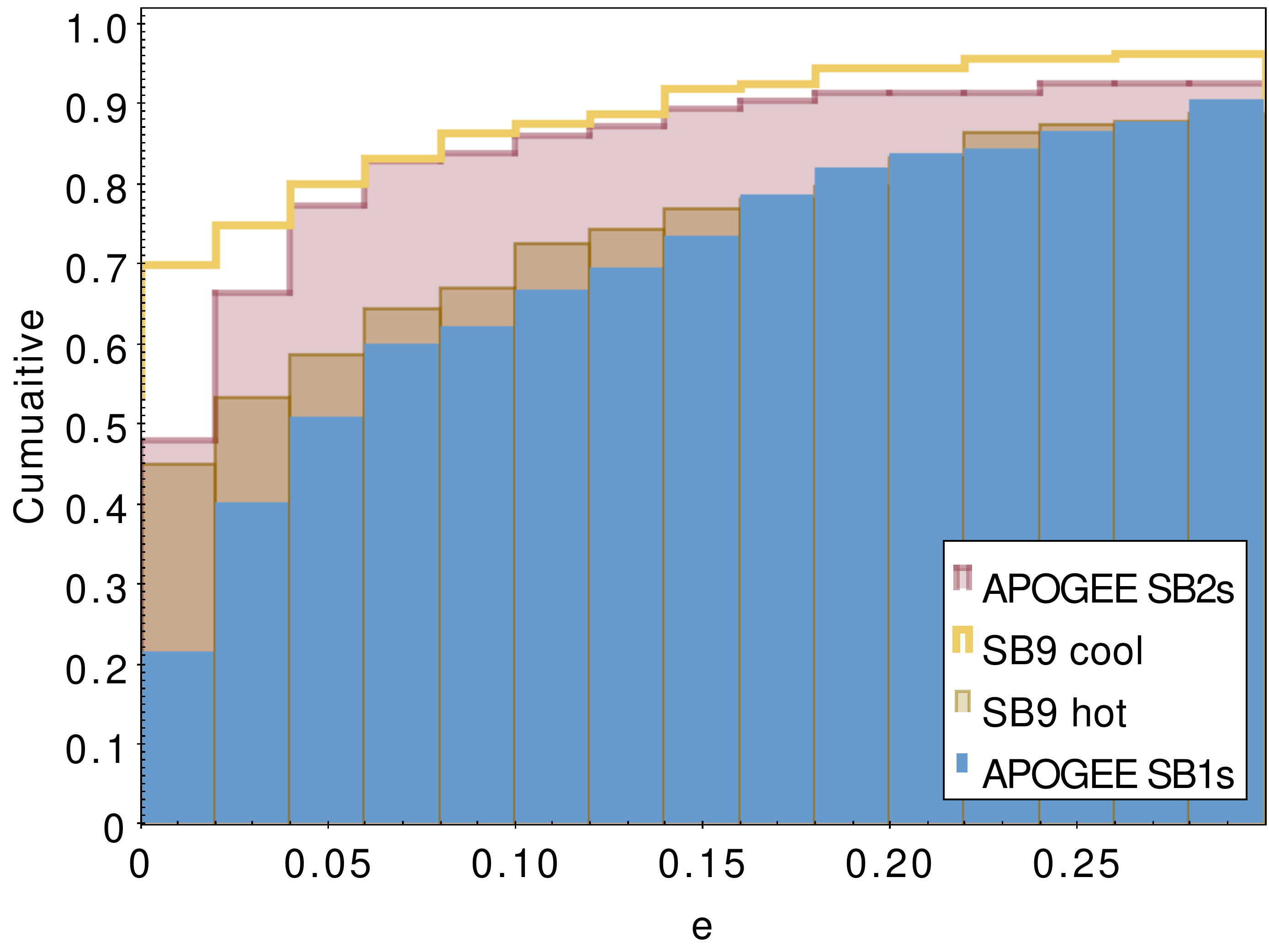}
\caption{Left: Period-eccentricity relationship of the fitted orbit. Note the plateau in eccentricity for systems with period $<10$ days due to the tidal circularization of evolved systems. The black lines shows the maximum eccentricity $e_{\mathrm{max}}(P)=1-(P/P_{\mathrm{min}})^{-2/3}$ from \citet{moe2017}, with $P_{\mathrm{min}}=$2, 0.5, and 0.2 days, which is the critical period at which the Roche lobe would overflow for equal mass binary with $M_1$=10, 2, and 0.5 \msun\ stars respectively. Right: Cumulative distribution of eccentricities for sources with $0.5<\log (P ~(\mathrm{day}))<1$ in APOGEE SB2s (this work), SB2s in the SB9 catalog with early type and late type primaries \citep{pourbaix2004}, and APOGEE SB1s \citep[both early type and late type primaries have a comparable distribution]{price-whelan2020}. Note that the cooler SB2s (both in this work and in SB9 catalog) tend to have fewer sources with higher $e$ due to more efficient circularization in comparison to hotter SB2s or SB1s with preferentially lower mass ratio.
\label{fig:pe}}
\end{figure*}

In the current (EDR3) data release, Gaia treats all sources as single when deriving their astrometric solutions. Motion of the photocenter due to unresolved multiplicity can affect the resulting astrometric fit, however, such that one of the metrics provided to evaluate the quality of the solution is the re-normalised unit weight error (RUWE). It is typically $<1.4$ for the sources for which the parallax is thought to be reliable \citep{lindegren2020}, and it is expected that many sources with RUWE$>1.4$ are likely multiples. Examining the distribution of RUWE, we find that sources with high RUWE are indeed more common among the spectroscopic binaries and higher order multiples we identify than they are in the full APOGEE sample. Furthermore, it is more common to find high RUWE among SB2s than it is among SB1s (Figure \ref{fig:ruwe}a). As the smallest velocity difference we are sensitive to with SB2s is $>$20\kms, and SB1s can be detected with smaller RV amplitude, it is reasonable to expect the orbital motion of SB2s to be more pronounced in astrometry as well. It should be noted that in $q\sim$1 systems, equal brightness of the stars in the pair may make it more difficult to detect multiplicity through RUWE \citep[as the center of light would be averaged out at the barycenter instead of being pulled by an individual star,][]{kervella2019}. Indeed, the fraction of sources with high RUWE tends to decrease as a function of the mass ratio (Figure \ref{fig:ruweq}). However, as SB2s tend to have more systems with high RUWE than SB1s, and SB2s also tend to have binaries with mass ratio towards unity, other factors that would make SB2s have a greater impact on astrometry must still dominate.

However, we note that most of the SB2s and higher order multiples do have RUWE$<1.4$ -- in most cases, the orbital separations of the systems identified here are too small to have a considerable effect on the Gaia astrometry. For the vast majority of spectroscopic binaries, including almost all systems for which we have been able to derive a complete spectroscopic orbit, the semi-major axis is too small for Gaia to resolve the orbital separation of the two stars, or the mass ratio is too close to unity to see the motion of the photocenter, and their resulting orbital motion cannot be detected when superimposed on the motion due to parallax and proper motion.

\citet{stassun2021} derived an expression for estimating the semi-major axis of a binary system based on RUWE for sources with RUWE between 1 \& 1.4. However, we find that in this sample, because spectroscopic binaries tend to have very short periods, this estimate only provides an upper bound for the semi-major axis derived from the orbital fitting. (Figure \ref{fig:ruwe}b).

\subsection{Binary Parameter Distributions and Current Observational Biases}

\subsubsection{Fraction of SB2s and Mass}

As the product of a targeted spectroscopic survey, the APOGEE dataset is subject to intentional selection effects that may bias the sample's SB2 fraction. Correcting for these selection effects to infer the intrinsic SB2 fraction as a function of stellar parameters is beyond the scope of this work, but as a first step in this direction, we measure the observed SB2 fraction ($F_{SB2}$) in our sample, relative to the total census of solar-type dwarfs observed by APOGEE. We exclude red giants from this analysis, as SB2s among them are rarely detected. We therefore limit the SB2 fraction analysis to sources with APOGEE Net \logg$>4$ dex, among both the SB2s and the underlying APOGEE comparison sample. Additionally, we limit this analysis to sources with \teff$<$6500 K, as both the surface gravities and RV values are less precise among hotter stars, making it more difficult to filter giants and subgiants and measure velocity amplitudes accurately. Similarly, the census of SB2s among hotter stars is highly incomplete since their faster rotational velocities result in broad CCF profiles, which are more difficult to reliably deconvolve into multiple components.

For the full sample of 5850 dwarfs, we measure an average $F_{SB2}$ of $\sim$3\%. As shown in Figure \ref{fig:ratio}, there is a strong trend in this fraction when we bin the sample according to the signal-to-noise (SNR) ratios of individual visit-level spectra, as noise in the data affects the strength of cross-correlation. As a result, our SB2 census is highly incomplete, and thus our observed $F_{SB2}$ is spuriously low, when limited to only low-SNR spectra. The $F_{SB2}$ rises steeply with SNR, however, and then stabilizes at $\sim$3.25\% for SNR$\gtrsim50$. However, low SNR data accounts for only 7\% of the total sample, so these sources do not add substantial biases to the $F_{SB2}$ we measure as functions of other stellar parameters. Similarly, there is no strong dependence in $F_{SB2}$ on the number of visits to a particular source, as identification of SB2s is performed on individual epochs. Although a larger number epochs may offer greater chance of catching a binary in a resolved state at least once, there is no particularly strong trend in the data. We note that the number of epochs over which a particular source is observed is not entirely independent of the stellar parameters (and as such, multiplicity), as repeat visits would be scheduled only to a particular type of targets.

We examine the raw, non-completeness corrected $F_{SB2}$ as a function of stellar temperature in Figure \ref{fig:ratio}, finding a distribution that is flat for \teff$<$6000K, but rises steeply for sources with \teff$>$6000K. Detection of SB2s in this study appears to be largely flat with respect to mass from M \& K dwarfs, with somewhat abrupt increase towards G dwarfs. This is similar to the temperature dependence of the multiplicity fraction of SB1s measured by \citet{price-whelan2020} and \citet{mazzola2020}. The similarity between these $F_{SB2}$, particularly in the \teff$<$6000K regime, is itself somewhat surprising: the completeness analysis by \citet{kounkel2019} indicates that SB2s should be easier to recover around higher mass primaries, such that an intrinsically flat multiplicity fraction would produce an observed $F_{SB2}$ that declines steadily and smoothly for \teff$<$6000K. The heightened sensitivity to SB2s with more massive primaries is partially due to the larger RV separation induced by more massive stars at comparable semi-major axes; furthermore, massive stars have been measured to have a higher intrinsic multiplicity fraction \citep{duchene2013}. However, we note that there may be additional selection biases in targeting strategy of the survey that may have resulted in a larger fraction of SB2s being targeted among the low mass stars due to their brightness compared to single stars.

\subsubsection{Eccentricity}

Dynamical evolution drives binaries towards more circular orbits over time. The timescale of this circularization process increases for longer orbital periods, such that short period systems circularize quickly, and long period systems can maintain substantial eccentricity for much of their stellar lifetimes. This behavior is apparent in Figure \ref{fig:pe}, which shows eccentricity as a function of period for the systems with fitted orbital solutions, and reveals that SB2s in our sample with $P<10$ days tend to have $e\sim0$, which is consistent with theories of orbital circularization due to tides. Other catalogs of short period binaries also find similar distributions in period-eccentricity space \citep[e.g.,][]{price-whelan2018a,raghavan2010,moe2017}.

Additional information about the circularization rate can be extracted from the power law slope $\eta$ of the eccentricity distribution $P_e\propto e^\eta$ for those sources with $P<10$ days, which circularize most efficiently. A steep eccentricity distribution ($\eta=-1$) indicates sources circularize rapidly, and the eccentricity distribution is more sharply peaked at $e=0$, whereas a flatter eccentricity distribution ($\eta=$0) indicates systems circularize more slowly, producing a shallower slope toward $e=$0. We find the slope of the eccentricity distribution of the APOGEE SB2 sample is $\eta\sim$-1. Comparing this distribution to other catalogs of multiples, this is in a good agreement for SB2s with low mass primaries located below the Kraft break in the SB9 catalog \citep{pourbaix2004}. However, it is steeper than the eccentricity distribution of hotter (\teff$>$6200K) SB2s in the \citet{pourbaix2004} catalog, which have $\eta\sim$-0.5, as tidal circularization is less efficient in these early type systems \citep[e.g.,][]{moe2017}. Interestingly, however, APOGEE SB1s in the gold sample from \citet{price-whelan2020} also show a shallower $e$ distribution with $\eta\sim$-0.5, without a considerable difference between early type and late type stars. SB1s generally have lower mass ratios than SB2s, and circularization timescales are 5 times longer for systems with $q\sim0.5$ than they are for $q\sim1$ \citep{zahn1977,hut1981}, which may explain why SB1s may have more sources with higher $e$ than SB2s detected within the same survey. Additionally, unlike SB2s, SB1s can be detected among red giants. This evolution may also dynamically affect the orbit.

Among the short period systems, however, there are still some short period SB2s with large $e$. Within their uncertainties, none exceed the critical $e$ threshold for the Roche lobe overflow for their mass, but they also do not have signatures of youth that might suggest they have not yet been circularized. As such, it is not entirely clear what may drive their higher eccentricities. For some of them, only a few RV data points drive their orbit toward a high $e$ solution; more observations may indicate that these RVs are inaccurate, and drive their orbits towards more expected $e$ values (we note that there is no significant difference in $e$ distribution as a function of the number of available epochs of RV measurements). However, some of these systems do have well-sampled RV curves, making their eccentric orbits more difficult to attribute to measurement errors. Such systems may include heartbeat stars \citep[e.g.,][]{shporer2016}, or an unseen tertiary companion which affects their orbit evolution. Cash, J. et al. (in prep) have identified 50 heartbeat stars in the APOGEE DR17 sample, of which 3 are SB2s presented here. Two of them (2M18515331+4043371, 2M19381285+4451011) have only a single APOGEE epoch. The third one, however, 2M19183945+4724013, does have 16 epochs and a full solved orbit, with $P$=2.7 days, and $e$=0.39. Detailed modeling, and follow-up observations, will help identify the mechanisms responsible for other unusually eccentric short-period orbits. 

\subsubsection{Fe/H}

Previous work has found good evidence that the multiplicity fraction of short period systems rises for lower metallicity primaries \citep[e.g., ][]{badenes2018, moe2019, mazzola2020}, but this signal may be suppressed at the shortest periods, where dynamical evolution plays a strong role in shaping the present-day multiplicity fraction \citep[e.g., ][]{hwang2020}. Cleanly detecting the metallicity dependence of the multiplicity fraction in a catalog of SB2s may also suffer from systematic effects due to 'veiling' by secondary light; in their analysis, \citet{traven2020} identified that SB2s in the GALAH sample have a lower Fe/H, on average by 0.2 dex, most likely due to bias introduced by the parameter extraction pipeline introduced by an SB2. 

We investigate the possibility of a metallicity bias in the SB2 detection fraction using stellar parameters produced by ASPCAP \citep{garcia-perez2016} and APOGEE Net \citep{olney2020}. Restricting the sample to dwarfs (\logg$>$4 dex) we find that SB2s appear to be systematically more metal poor than single stars, although the effect is significantly more pronounced when adopting [Fe/H] values from APOGEE Net than from ASPCAP, where the metallicity offset between SB2s and single stars is slight. It is possible that other APOGEE Net parameters are similarly affected by the presence of secondary light. \citet{olney2020} searched for, but did not find, a systematic shift in \logg\ between single and multiple pre-main sequence stars; examining APOGEE Net's \logg\ values for our SB2 sample and the comparison sample of main sequence dwarfs, we see evidence for a difference in the median \logg\ of $\sim$0.12 dex. Curiously, although there is a comparable magnitude of discrepancy in \logg\ in ASPCAP, the trend is reversed. No clear difference is apparent in either catalog's \teff\ distributions.

\begin{figure*}
\epsscale{1.1}
\plottwo{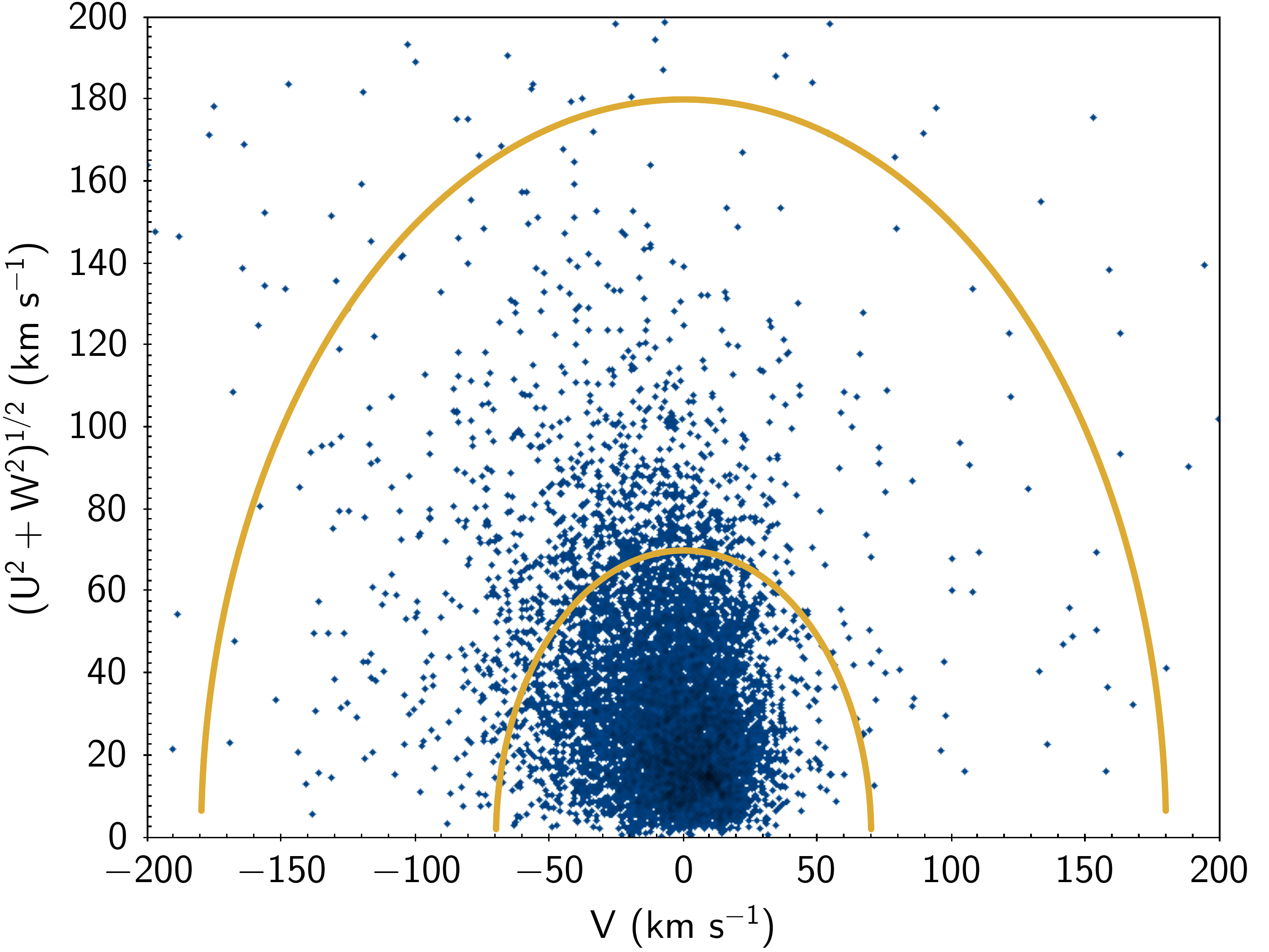}{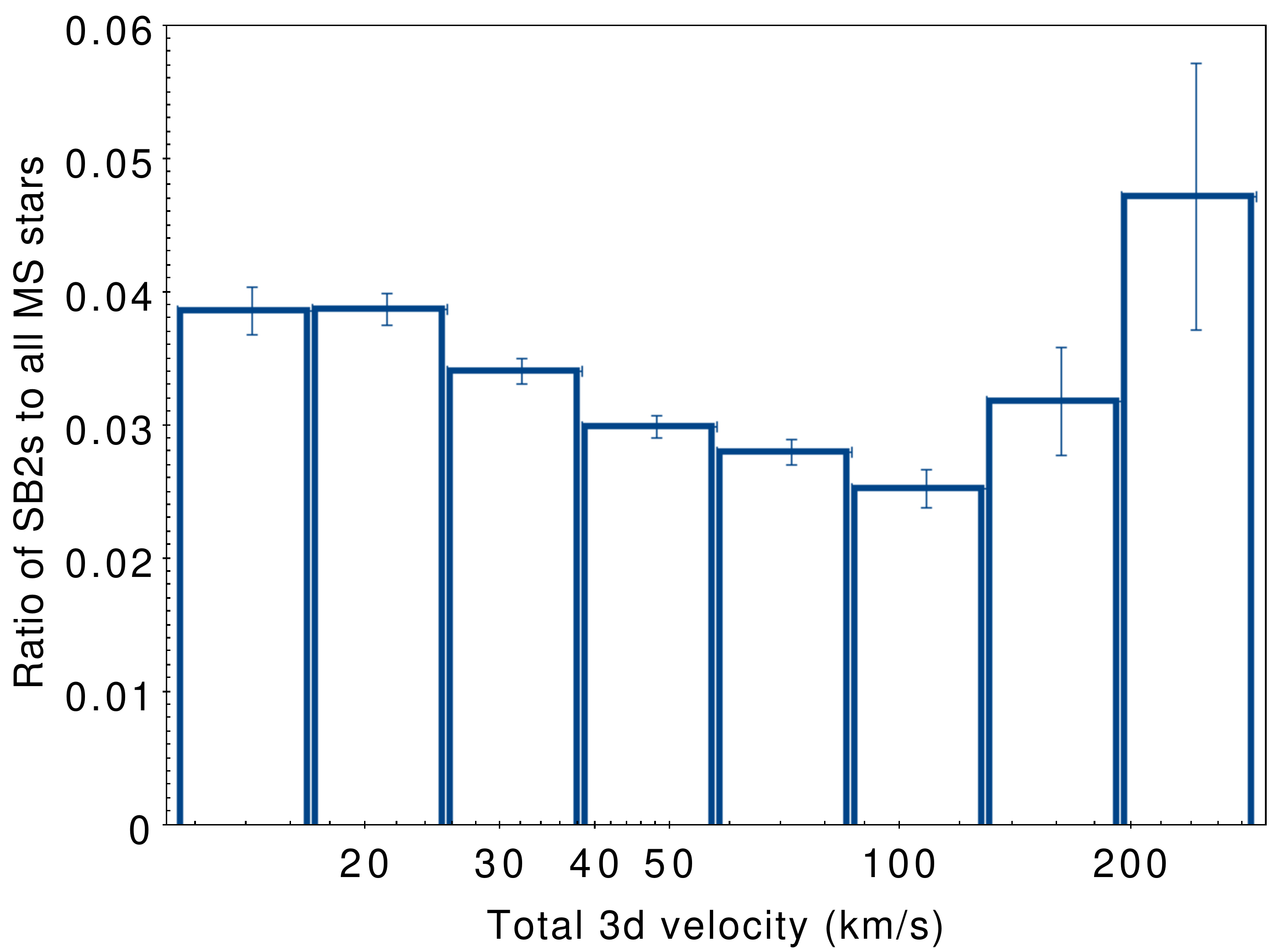}
\caption{Left: Toomre diagram, showing UVW kinematics of SB2s. The two semi-circles are placed at V$_{tot}$=70 and 180 \kms, roughly separating the typical kinematics of the thin disk, thick disk, and the halo. \citep[e.g.,][]{bensby2005, reddy2006, nissen2009}. Right: Fraction of SB2s relative to the full APOGEE sample of main sequence stars, as a function of V$_{tot}$.
\label{fig:toomre}}
\end{figure*}

To test if these parameter offsets are intrinsic/astrophysical in nature, or introduced by detection/measurement bias, we investigate if the metallicity offset in our SB2 catalog is consistent with known correlations between metallicity and kinematically defined Galactic populations. That is, we use kinematics-based membership of the thin disk, thick disk, and the halo as a proxy for metallicity, and search for consistent offsets in the SB2 fraction of these populations. Examining the kinematics of the sample using the Toomre diagram (Figure \ref{fig:toomre}), it appears that the bulk of the sample, $\sim$80\% consists of the thin disk stars. 

Curiously, the $F_{SB2}$ appears to decrease with the total magnitude of 3D velocity of the star, $V_{tot}$, for thin and thick disk stars. However, it increases for halo stars.

\begin{figure*}
\epsscale{1.1}
\plottwo{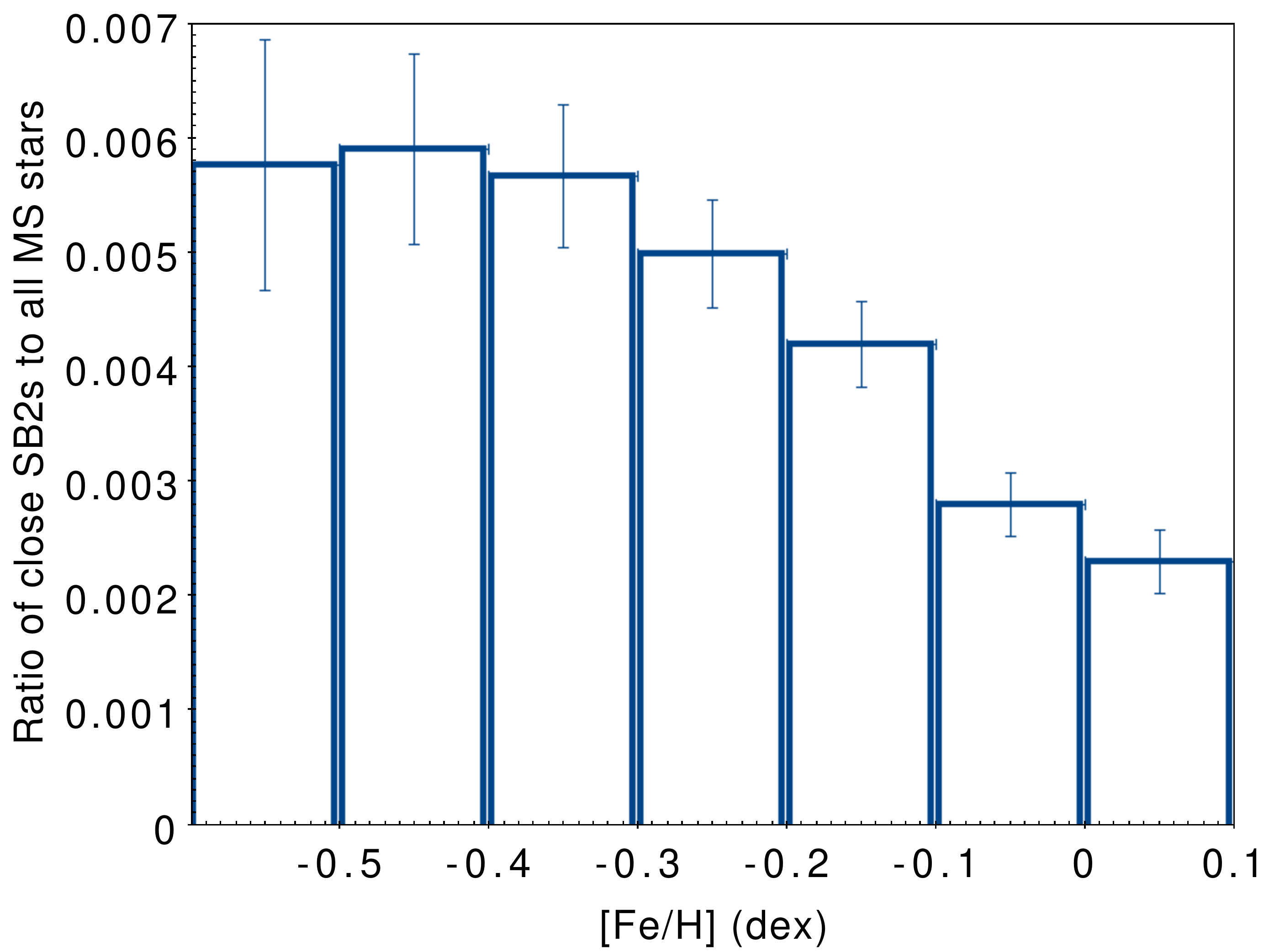}{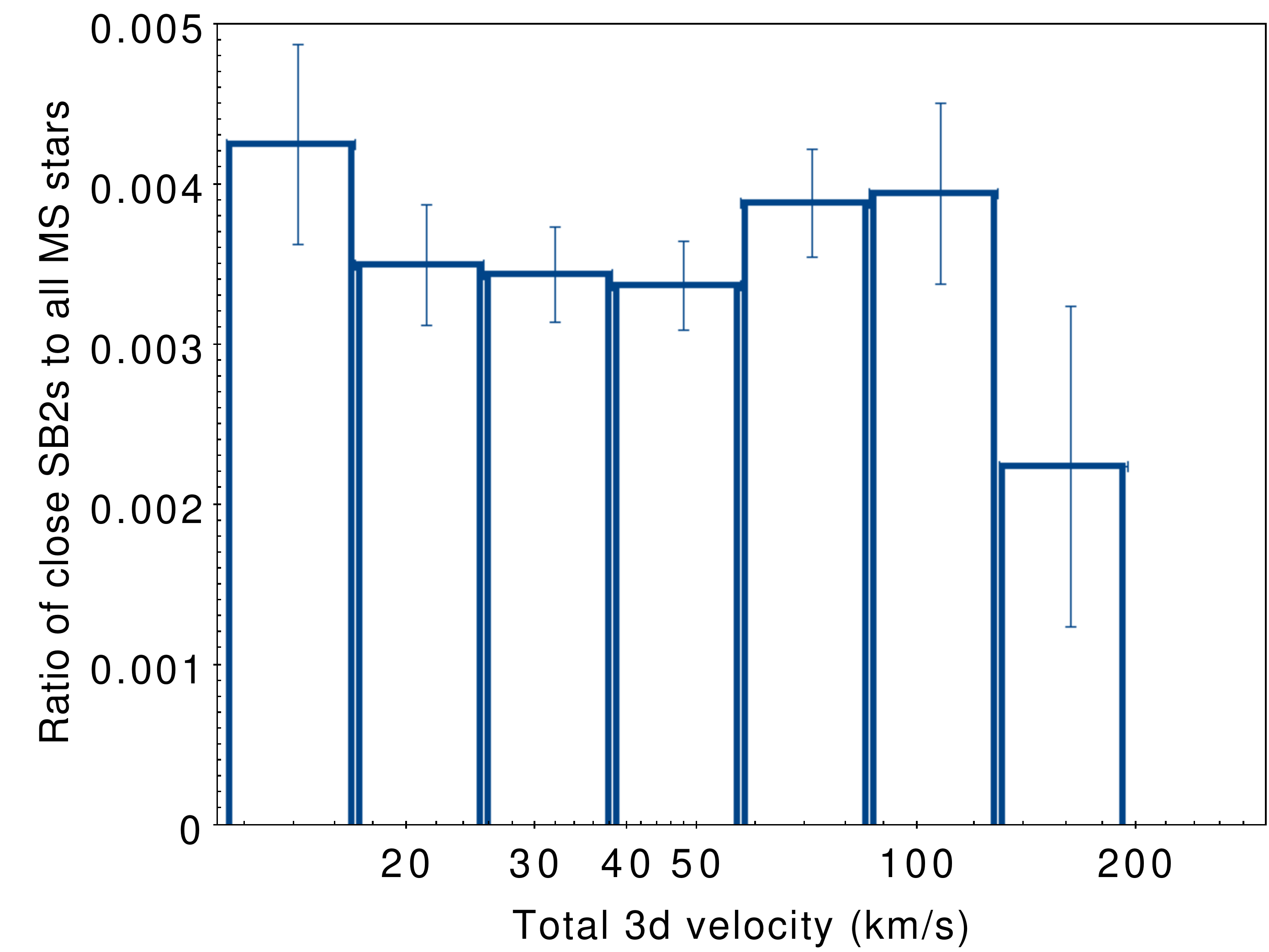}
\caption{Fraction of SB2s with $q>0.8$ and the maximum separation between two peaks of $>60$ \kms to the full APOGEE sample of main sequence stars. Left: as a function of Fe/H from ASPCAP. Right: as a function of V$_{tot}$.
\label{fig:toomre1}}
\end{figure*}

One explanation for such a trend is that metal poor stars tend to have weaker lines in their spectra. This may lead to a weaker signal in the CCF, especially in the weakly resolved systems with large disparity in their masses. This could lead to a suppression in the observed $F_{SB2}$ of metal poor-stars, although at very low Fe/H, the intrinsically high $F_{SB2}$ of the metal-poor stars is able to overcome this observational bias. Restricting the sample of SB2s to only systems with $q>0.8$ and RV separation between two stars of $>60$ \kms\ (the parameter space that is most likely to be complete across stars of all Fe/H), we recover the trend of higher $F_{SB2}$ in low Fe/H systems (Figure \ref{fig:toomre1}). Similarly, the observed $F_{SB2}$ as a function of $V_{tot}$ for these equal mass and well resolved systems no longer suppresses the $F_{SB2}$ of thick disk stars (with $V_{tot} \sim$100 \kms).

\begin{figure*}
\epsscale{1.1}
\plottwo{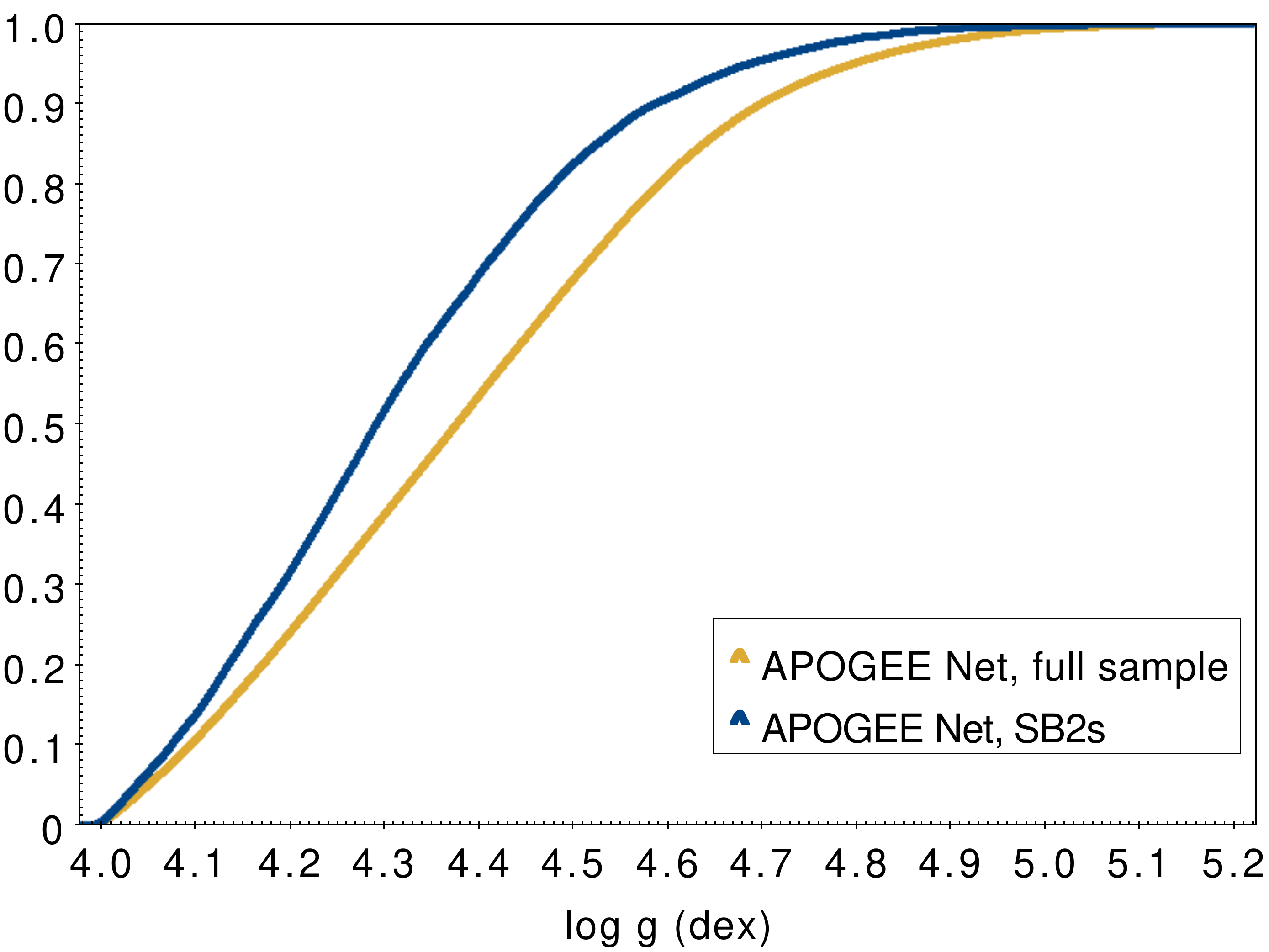}{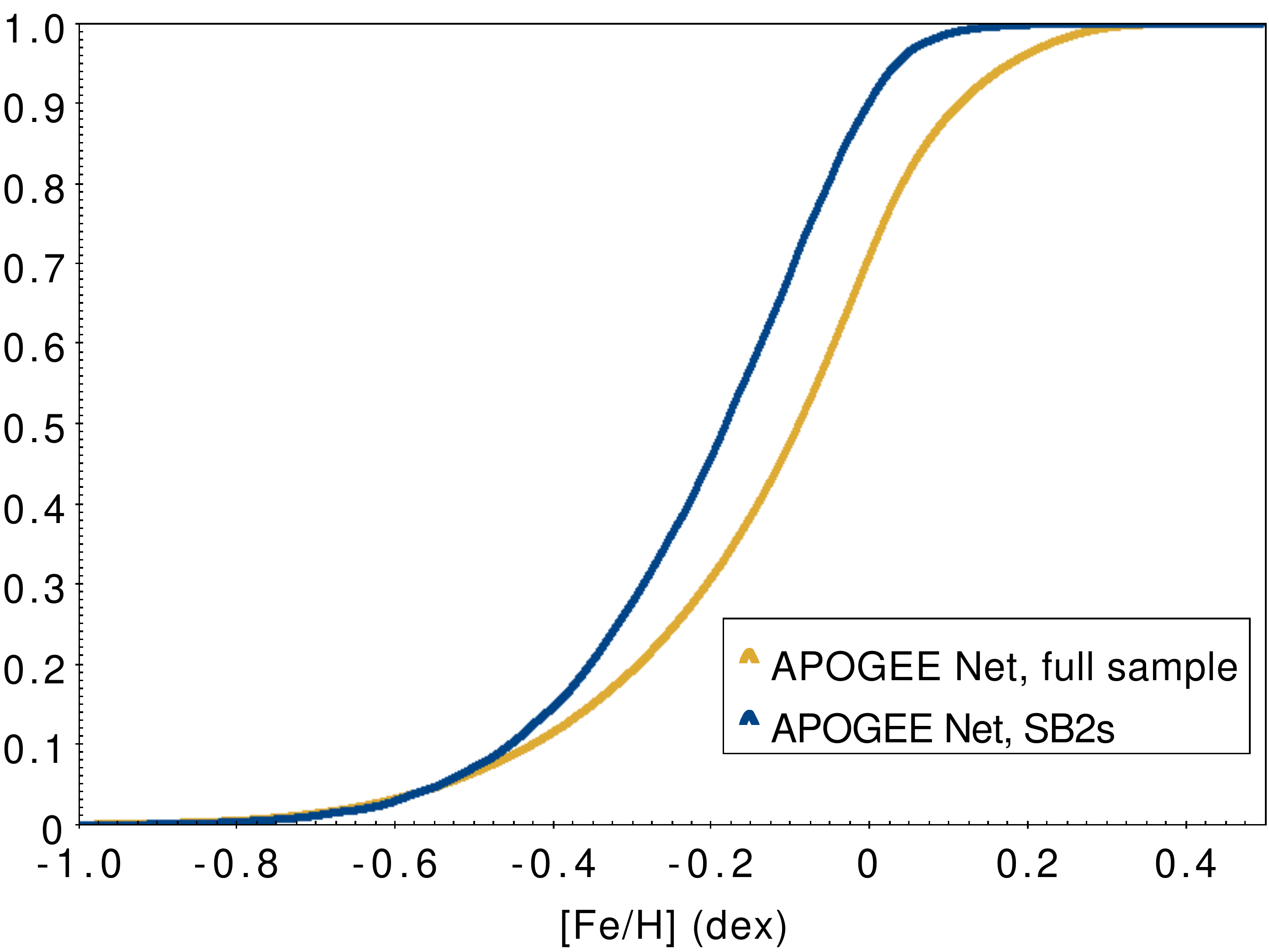}
\plottwo{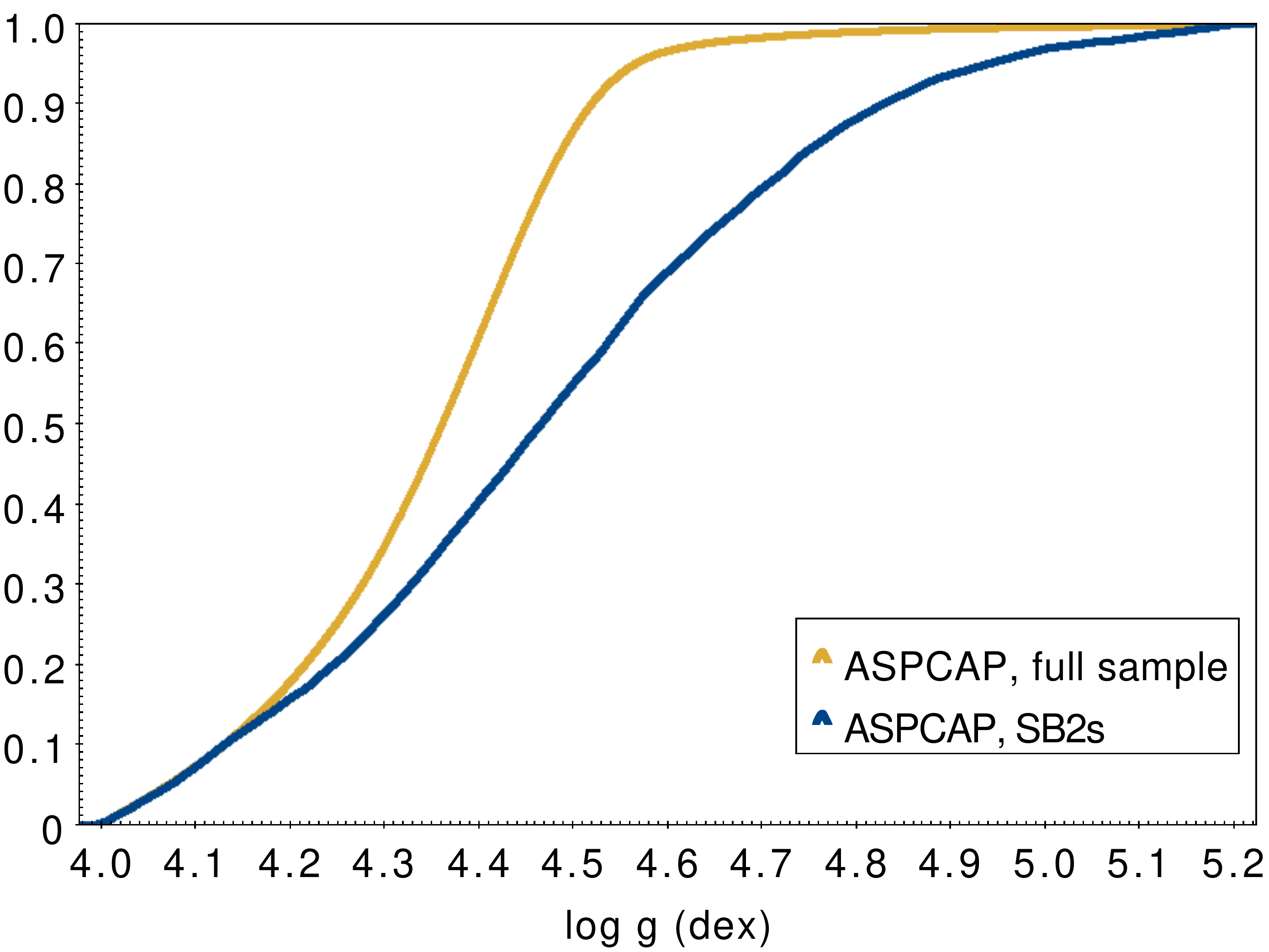}{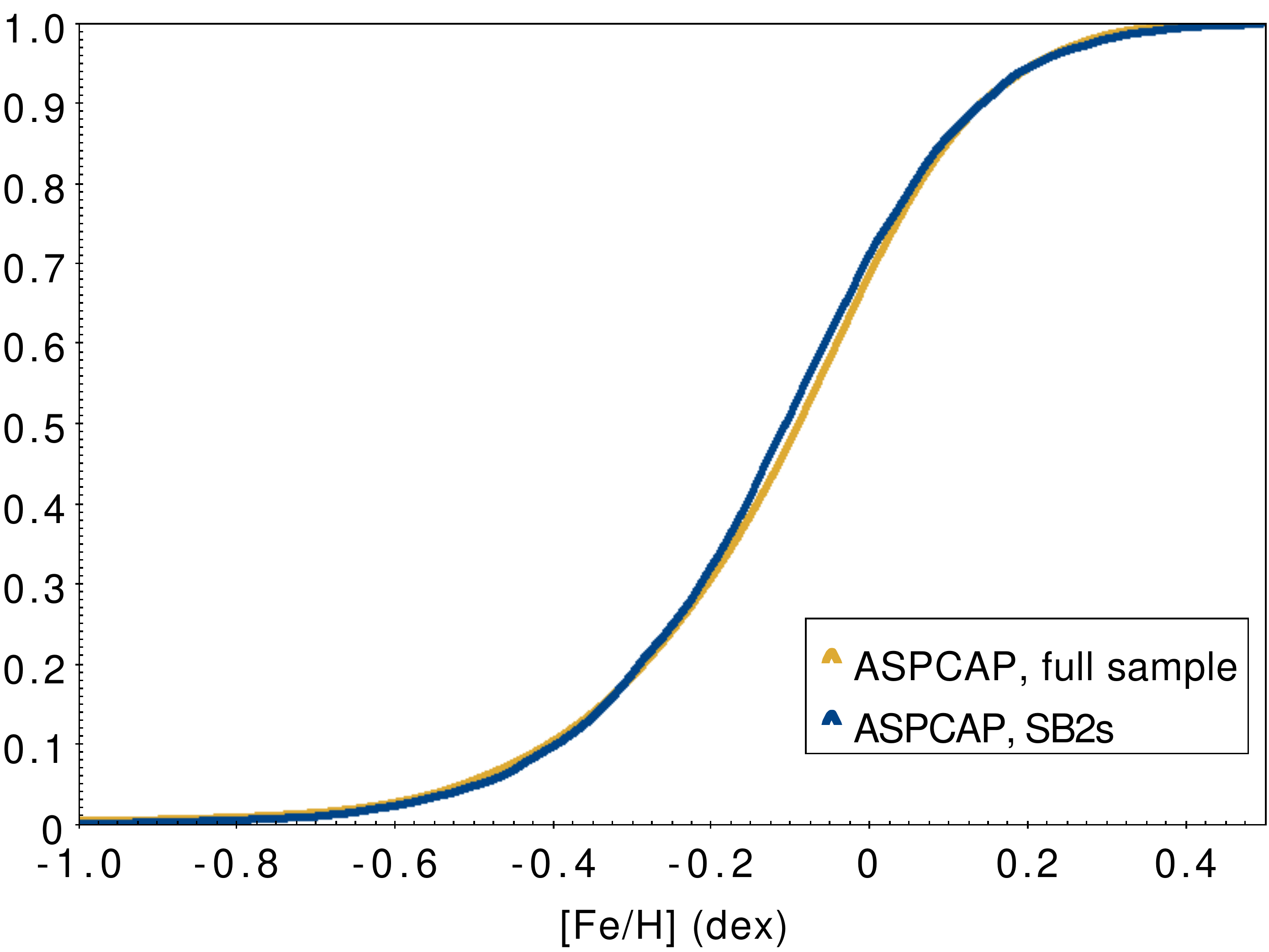}
\caption{Cumulative distribution of stellar parameters derived by ASPCAP \citep[Holtzmann, J. et al. in prep]{garcia-perez2016} and APOGEE Net \citep{olney2020} for all the stars in the APOGEE sample (with \logg$>$4 dex and \teff$<6500$ K) in yellow, vs their distribution within the sample of SB2s satisfying the same cuts in blue. Typical uncertainties are 0.02 dex in Fe/H and 0.05 dex in \logg.
\label{fig:feh}}
\end{figure*}

\section{Conclusions}\label{sec:concl}

We developed an autonomous pipeline to identify double-lined and higher order spectroscipic binaries in the APOGEE spectra through Gaussian deconvolution. We visually examined all of the fitted components in the intermediate DR16+ data to improve the fit and remove spurious detections. Due to an incompatible data model, DR17 data have only been processed by the autonomous pipeline, without being merged into the visually vetted catalog.

In total, we identify 7273 SB2s (some of which may have undetected third component based on unusual mass ratios) 813 SB3s, and 19 SB4s. At this time, we are able to derive complete orbits for 325 of these SB2s. Furthermore, we examine TESS light curves, identifying variable stars, of which 369 appear to be detatched eclipsing binaries. A complete modeling of both radial velocities and the light curves for such systems would make it possible to fully characterize both the masses and the radii of the individual stars.

Unlike SB1s, SB2s can be detected as such in just a single epoch, however, there is a minimum signal-to-noise ratio requirement of $\sim$50 to ensure optimal detection. At the resolution of APOGEE, the radial velocities of two components need to be separated by at least 20 \kms\ to be able to resolve them clearly.

The derived orbital parameters are consistent with the previously derived distributions. In particular, sources with period $<$10 days tend to be circularized. This circularization is stronger in cool SB2s in comparison to hotter SB2s or to SB1s.

The SB2 fraction appears to be largely flat with mass for slowly rotating stars below the Kraft break. SB2s are strongly biased towards the dwarfs due to the necessity of both stars having comparable flux - true SB2s among giants are rare. However, there are line of sight coincidence systems that can capture the spectrum of two or more unrelated stars - among them, the red giants are more common. Trends with metallicity are difficult to observe among SB2s due to various observational biases -- in part because of it, SB2s appear to be less common among thick disk stars compared to the stars in the thin disk or in the halo. Restricting the sample to only well resolved systems is able to recover previously observed trends of increasing $F_{SB2}$ of close companions in metal poor systems.

\appendix
\restartappendixnumbering

\section{DR17}\label{sec:appendix}

The data presented in this paper span the period until the shutdown of APOGEE in April/May 2020, using the intermediate data release made available to the collaboration. However, the data reduction pipeline and data model has changed since then for DR17 data release.

The two reductions, and the CCFs produced by them, are not directly compatible with one another. Although SB2s with widely separated RVs of both components tend to be recovered in both reductions, this is not always the case when they appear to be more blended in some epochs. A particular peak in CCF might appear as a composite of two components in one data release, and as a single component in the other release, and vice versa. There does not appear to be a systematic bias between which pipeline performs better in this regard.

Furthermore, the APOGEE data processing pipeline may fail to extract a primary RV in a spectrum - in this case, no apVisit file is produced for that epoch in the data release. DR17 tends to have a more stable performance than the intermediate release, and it includes more successfully processed visits in the same range of time. However, there are some visits that were included in the intermediate data release but not in DR17.

Because much of the analysis in the paper relied on visual inspection and validation of the data, both on a per epoch and on a per system level, such differences in the data make the inclusion of the data that are part of DR17 but not the intermediate release difficult after the bulk of the analysis was already underway. 

Therefore, in addition to the curated catalog of binary velocities included in Table \ref{tab:system} that is based on the intermediate data release, we apply \texttt{apogeesb2} on the DR17, and present the measured velocities as is in Table \ref{tab:dr17}.

We note that DR17 natively includes a version of an SB2 search algorithm roughly based on \citet{kounkel2019}, however, it has lower completeness, and a higher contamination rate, most notably among higher mass and lower mass stars.

\begin{deluxetable*}{cccccccccc}
\tabletypesize{\scriptsize}
\tablewidth{0pt}
\tablecaption{Parameters of the CCF components extracted by the pipeline in APOGEE DR17 data\label{tab:dr17}}
\tablehead{
\colhead{APOGEE} &\colhead{$\alpha$} &\colhead{$\delta$} &\colhead{HJD} &\colhead{$v_1$} & \colhead{FWHM$_1$} & \colhead{Amp$_1$}& \colhead{Flag$_1$\tablenotemark{$^a$}}& \colhead{N}&\colhead{DR16}\\
\colhead{ID} &\colhead{(deg.)} &\colhead{(deg.)} &\colhead{(day)} &\colhead{(\kms)} & \colhead{} & \colhead{}&\colhead{}&\colhead{comp}& \colhead{comparison\tablenotemark{$^b$}}
}
\startdata
2M00004521-7219055 & 0.188401 & -72.318222 & 59154.57 & 10.9$\pm$4.2 & 25.6$\pm$6.4 & 0.31$\pm$0.05 & 4&2& New obs \\
\enddata
\tablenotetext{}{Only a portion shown here. Full table (with up to 4 components) is available in an electronic form.}
\tablenotetext{a}{1=Failed FWHM/amplitude test; could be noise. 2=Failed the symmetry test; could be falsely multiply deconvolved. 3=Blended asymmetric peaks; generally a robust detection of a companion, but may be a signature of spots in strongly magnetic stars. 4=The primary peaks in all stars, and the secondary peaks of bona fide SB2s that are not blended with the primary}
\tablenotetext{b}{New obs=Epoch not a part of Table \ref{tab:epoch}; New comp=Epoch in Table \ref{tab:epoch} is unresolved with a single component; component with Flag$\geq$3 is detected in DR17}
\end{deluxetable*}


\software{GaussPy \citep{lindner2015}, The Joker \citep{price-whelan2017}, rvfit \citep{rvfit}, eleanor \citep{feinstein2019}, TOPCAT \citep{topcat}, apogeesb2 \citep{apogeesb2}}

\begin{acknowledgments}

M.K. and K.C. acknowledge support provided by the NSF through grant AST-1449476, and from the Research Corporation via a Time Domain Astrophysics Scialog award (\#24217). M.K. acknowledges support from NASA ADAP grant 80NSSC19K0591.
M.M. and K.M.K. acknowledge financial support from NASA under grant ATP-170070.
CRZ acknowledges support from projects UNAM-DGAPA-PAPIIT 112620 and CONACYT CB2018 A1-S-9754, Mexico.
JH acknowledges support from projects UNAM-DGAPA-PAPIIT IA-102921 and CONACYT-CF86372.
DAGH acknowledges support from the State Research Agency (AEI) of the Spanish Ministry of Science, Innovation and Universities (MCIU) and the European Regional Development Fund (FEDER) under grant AYA2017-88254-P.
JS acknowledges from the CONACYT by fellowship support in the Posgrado en Astrof\'{i}sica graduate program at Instituto de Astronom\'{i}a, UNAM.
P.M.F. acknowledges support provided by the NSF through grant AST-1715662.
T.C.B. acknowledges partial support from grant PHY 14-30152 (Physics Frontier Center/JINA-CEE), awarded by the U.S. National Science Foundation. RHB thanks support from ANID FONDECYT Regular Project No. 1211903.

Funding for the Sloan Digital Sky Survey IV has been provided by the Alfred P. Sloan Foundation, the U.S. Department of Energy Office of Science, and the Participating Institutions. SDSS-IV acknowledges
support and resources from the Center for High-Performance Computing at
the University of Utah. The SDSS web site is www.sdss.org.
SDSS-IV is managed by the Astrophysical Research Consortium for the 
Participating Institutions of the SDSS Collaboration including the 
Brazilian Participation Group, the Carnegie Institution for Science, 
Carnegie Mellon University, the Chilean Participation Group, the French Participation Group, Harvard-Smithsonian Center for Astrophysics, 
Instituto de Astrof\'isica de Canarias, The Johns Hopkins University, 
Kavli Institute for the Physics and Mathematics of the Universe (IPMU) / 
University of Tokyo, Lawrence Berkeley National Laboratory, 
Leibniz Institut f\"ur Astrophysik Potsdam (AIP), 
Max-Planck-Institut f\"ur Astronomie (MPIA Heidelberg), 
Max-Planck-Institut f\"ur Astrophysik (MPA Garching), 
Max-Planck-Institut f\"ur Extraterrestrische Physik (MPE), 
National Astronomical Observatories of China, New Mexico State University, 
New York University, University of Notre Dame, 
Observat\'ario Nacional / MCTI, The Ohio State University, 
Pennsylvania State University, Shanghai Astronomical Observatory, 
United Kingdom Participation Group,
Universidad Nacional Aut\'onoma de M\'exico, University of Arizona, 
University of Colorado Boulder, University of Oxford, University of Portsmouth, 
University of Utah, University of Virginia, University of Washington, University of Wisconsin, 
Vanderbilt University, and Yale University.
This work has made use of data from the European Space Agency (ESA)
mission {\it Gaia} (\url{https://www.cosmos.esa.int/gaia}), processed by
the {\it Gaia} Data Processing and Analysis Consortium (DPAC,
\url{https://www.cosmos.esa.int/web/gaia/dpac/consortium}). Funding
for the DPAC has been provided by national institutions, in particular
the institutions participating in the {\it Gaia} Multilateral Agreement.
\end{acknowledgments}

\bibliographystyle{aasjournal.bst}
\bibliography{apbin.bbl}

\end{document}